# REDSHIFT-INDEPENDENT DISTANCES IN THE NASA/IPAC EXTRAGALACTIC DATABASE: METHODOLOGY, CONTENT, AND USE OF NED-D

Ian Steer[1], Barry F. Madore[2,3], Joseph M. Mazzarella[3], Marion Schmitz[3], Harold G. Corwin, Jr.[4], Ben H. P. Chan[3], Rick Ebert[3], George Helou[3], Kay Baker[3], Xi Chen[3], Cren Frayer[3], Jeff Jacobson[3], Tak Lo[3], Patrick Ogle[3], Olga Pevunova[3], and Scott Terek[3]

1. 268 Adelaide St. E., Ste. 188, Toronto, ON M5R 2G2, Canada; steer@bell.net

2. Observatories of the Carnegie Institution for Science, 813 Santa Barbara St., Pasadena, CA 91101, USA; barry@obs.carnegiescience.edu

3. Infrared Processing & Analysis Center, MS 100-22, California Institute of Technology, Pasadena, CA 91125, USA

4. 68 The Common, Williamsville, NY 14221, USA

ABSTRACT

Estimates of galaxy distances based on indicators that are independent of cosmological redshift are fundamental to astrophysics. Researchers use them to establish the extragalactic distance scale, to underpin estimates of the Hubble constant, and to study peculiar velocities induced by gravitational attractions that perturb the motions of galaxies with respect to the "Hubble flow" of universal expansion. In 2006 the NASA/IPAC Extragalactic Database (NED) began making available a comprehensive compilation of redshift-independent extragalactic distance estimates. A decade later, this compendium of distances (NED-D) now contains more than 100,000 individual estimates based on primary and secondary indicators, available for more than 28,000 galaxies, and compiled from over 2,000 references in the refereed astronomical literature. This article describes the methodology, content, and use of NED-D, and addresses challenges to be overcome in compiling such distances. Currently, 75 different distance indicators are in use. We include a figure that facilitates comparison of the indicators with significant numbers of estimates in terms of the minimum, 25th percentile, median, 75th percentile, and maximum distances spanned. Brief descriptions of the indicators, including examples of their use in the database, are given in an Appendix.

1. INTRODUCTION

Distances to nearby galaxies, based on stellar distance indicators such as Cepheid variable stars and luminous blue supergiants, were key in establishing the scale size of the universe and calibrating its expansion rate (Hubble 1926, 1929). Knowledge of an extragalactic object's distance allows one to move beyond apparent properties (such as angular sizes and apparent magnitudes, scaling by various powers of the distance) to physical quantities such as metric sizes, true volumes, energy densities and absolute luminosities across the electromagnetic spectrum, where models can interface directly with properly scaled data. However, individual distances are hard to determine even for the nearest of galaxies.

Distances to more remote objects are, in principle, more easily estimated using the redshift-distance relation or Hubble law, $D = v/H$, which gives the distance, D once the galaxy's recessional velocity, v and the Hubble parameter, H have been independently measured and calibrated. Redshifts are now available for millions of galaxies, from which redshift-based (cosmological Hubble flow) distances can be calculated, only to the degree that the expansion is assumed to be smooth, (unperturbed) homogenous and isotropic. That is only asymptotically true on the very largest scales, however. In fact, the measured redshift of any galaxy can have "peculiar" velocity components with respect to the overall flow; moreover, the velocity of galaxies relative to one another due to gravitational interactions within groups and clusters can be in excess of 1,000 km s$^{-1}$.

Redshift-independent distances, in contrast to redshift-based estimates, are based on indicators that are independent of galaxy recessional velocity. They are crucial to cosmology for at least two very important reasons. First, they set the size scale for the local universe, not only by providing metric distances to the nearest individual galaxies themselves, but also by tying in to other distance indicators and calibrating them in individual galaxies and/or in clusters of galaxies. Second, redshift-independent distances in the local universe set the foundations for determination of the scale size of the universe as a whole as well as of the Hubble constant and other cosmological parameters.

Redshift-independent distances for galaxies come in various flavors. There are so-called primary, secondary (and even tertiary) distance indicators. Primary distance indicators include standard candles such as Cepheids and Type Ia supernovae, and standard rulers such as megamasers. Primary indicators provide distances with precision typically better than 10%. In-depth reviews are available (see for example Ferrarese et al. 2000, Freedman & Madore 2010, de Grijs et al. 2014a, 2014b, 2015, and references therein). Distances with quoted precisions of better than 5% are available for some Local Group galaxies (e.g., Riess, Fliri, & Valls-Gabaud 2012).

To date the most statistically precise and systematically accurate distance for a galaxy beyond the Local Group has a total error of 4% (Humphreys et al. 2013). Secondary indicators include the Tully-Fisher and Fundamental Plane relations and provide distances with precision typically around 20%. In-depth reviews of distance indicators in general, including secondary indicators, are available (Tully et al. 2009, 2013, 2016, and references therein). For quick reference, brief descriptions of the indicators currently in use are provided in an Appendix following this article.

Enumerating and evaluating the published redshift-independent distances for a galaxy is often one of the first steps taken by researchers starting a study of any nearby galaxy. Ensuring that any given tabulation is complete and up to date is increasingly challenging however. First, many redshift-independent distances are published in papers that do not include the key words "distance" in their abstracts. So researchers cannot narrow their searches by combining the keywords for their galaxy of interest with the keyword "distance". Instead, all papers referencing galaxies of interest must be searched. Second, although less abundant than redshift distances, there are nevertheless many thousands of redshift-independent distance estimates published for thousands of galaxies in nearly a hundred references annually. Third, there are now at least 75 different redshift-independent distance indicators in use, compared to around a dozen before the precision era of extragalactic research (i.e., based on CCD and/or space telescope observations). Decisions regarding which indicators to include or exclude from any given analysis are often subjective, where the pros and cons of any one indicator over another are still hotly debated. Fourth, for many of the indicators and even for estimates using the same indicator, direct inter-comparison between published estimates requires some experience and careful reading of the primary literature. Estimates are often (subtly) based on different extragalactic distance scales. These can involve different distance scale zero points or values of the Hubble constant (for review see Freedman & Madore 2010), or different cosmological distance measures, including the linear or proper motion distance, the luminosity distance, and the angular diameter distance (for review see Hogg 1999).

Because of their value to researchers, numerous compilations of redshift-independent distance estimates have been made in recent years that are noteworthy. They include the first large compilation of primary distance estimates published in the precision era, compiled for the National Aeronautics and Space Administration (NASA) Hubble Space Telescope (HST) Key Project on the extragalactic distance scale by Ferrarese et al. (2000). Their paper featured 200 primary estimates for 100 galaxies. The Updated Nearby Galaxy Catalog, a compilation by Karachentsev et al. (2013), features mainly primary estimates for a thousand galaxies. The Hyper Lyon-Meudon Extragalactic Database (HyperLeda) catalogue provides nearly seven thousand distance estimates for more than two thousand galaxies found in 430 references (Makarov 2014). Finally, the Extragalactic Distances Database (EDD), an ongoing project by Tully et al. (2009, 2013, 2016), currently includes primary estimates for nearly one thousand galaxies and secondary estimates for many thousands more and growing.

Rapid growth in both the types and numbers of distance estimates published annually prompted members of the joint NASA/Infrared Processing and Analysis Center (IPAC) Extragalactic Database (NED) team a decade ago to introduce a new service, "NED Distances", hereafter referred to as NED-D. NED itself was created to provide researchers with easy access to extragalactic data integrated from thousands of peer-reviewed, published articles and catalogs via a single online service (Helou et al. 1990). In the following quarter century, NED has evolved and expanded to include many new data types and enable powerful science queries, using data from very large sky surveys and many derived quantities that explicitly depend on reliable distances (Mazzarella et al. 2001, 2007, 2014, Schmitz et al. 2011, Ogle et al. 2015). Initially, for redshift-independent extragalactic distances, a six-month pilot project was undertaken to tabulate distance estimates based exclusively on primary indicators published up to 2005, as briefly described by Mazzarella et al. (2007). What began as a snapshot review of the literature has grown into a baseline service provided by NED, aimed at maintaining the most complete and up-to-date compilation of all redshift-independent distance estimates published from 1980 to present. As of January 2016, over 100,000 separate distance estimates for more than 28,000 individual galaxies compiled from over 2,000 references are available in NED-D.

The volume of data permits programmed rule-based filtering to uniformly construct galaxy samples based on a variety of distance indicators, both for comparison of distance indicators and for other global analyses. This also permits NED users to select or filter subsets of interest. While choices regarding which estimates and/or indicators to include in any given analysis can be controversial, having a complete and comprehensive compilation of all available data for all interested parties is a requisite first step.

A major difference between other tabulations and the approach taken in Ferrarese et al. (2000) and in NED-D involves inclusion versus exclusion of data. Whereas distances based on multiple indicators as well as multiple estimates based on individual indicators are included in Ferrarese et al. (2000), other tabulations serve the function of providing a highly vetted subset of available distance estimates. Typically, one "best estimate distance" per galaxy is provided, usually based on the most recent estimate employing what the author considers to be the most precise indicator at the time of writing. Cepheids-based and Tip of the Red Giant Branch (TRGB)-based estimates for example, are often preferred because they are considered more precise than other indicators, including Surface Brightness Fluctuations (SBF) and the Planetary Nebulae Luminosity Function (PNLF). Comprehensive lists of multiple estimates based on a variety of indicators exist, but generally for individual galaxies, most notably the thorough tabulations for the Large Magellanic Cloud (LMC), Small Magellanic Cloud (SMC), and Messier 31, by de Grijs et al. (2014a, 2014b, 2015). They are investigating distance estimates for the LMC and other Local Group galaxies because of suggestions that recent values cluster around mean values more tightly than their individually quoted statistical errors would predict. The potential for a "band-wagon effect" to impact distance estimates to the LMC was first suggested by Schaefer (2008).

NED as a whole exists to support individual scientists, space missions, and ground-based observatories in planning, interpretation, and publication of research on galaxies, extragalactic distances, and cosmology. NED-D is being maintained as part of NED core activities, with updates and upgrades made on a regular basis. In this article, we describe the construction of NED-D, characterize its content and discuss planned future directions in its growth and evolution. The methods used in compiling NED-D are described in Section 2, and its formatting and accessibility are described in Section 3. Brief discussions and visualizations of the growth in distance data, statistical distributions of the indicators, and research activity behind these data, are given in Section 4. We report on NED-D's status and current plans for its future in Section 5, focusing on improving the accessibility, usability and scientific impact of redshift-independent distances. We summarize NED-D's first ten years and future prospects in Section 6, and include brief descriptions of the distance indicators currently in use in an Appendix, along with examples of their use in the database.

2. HOW NED-D IS COMPILED

Mining redshift-independent distance estimates differs from mining other data for NED. Most NED data involves object apparent properties, e.g., apparent positions, apparent magnitudes, apparent diameters, etc. Such data involve measurements that are model independent and result from direct observations. Redshift-independent distance estimates on the other hand, require modeling of the relations between apparent and absolute properties. Models in turn require choices that impact all estimate-based data. Factors impacting redshift-independent distance estimates, for example, include assumptions regarding the distance scale. In the case of standard candles, numerous corrections that account for reddening, age/metallicity, crowding, and more also need consideration.

Mining eligible distance estimates versus conventional measurement data is challenging in another way. Measurements, whether of redshifts or positions or other data, are generally published solo, i.e., without old, auxiliary and/or ancillary data. In contrast, new distance estimates by virtue of their model dependence are almost always published along with multiple values, for example the apparent distance versus the reddening corrected distance to galaxy NGC 1365 based on Cepheids (e.g., Leonard et al. 2003). Further, most new distance estimates are accompanied by numerous examples of previously published estimates for comparison. Detailed understanding of extragalactic distance scale research is required to properly identify, extract, and represent this information in a database.

For NED-D, an eligible distance estimate is defined as an originally published estimate of the distance to an extragalactic object, such as galaxy Messier 106 or quasar 3C 279, based on a redshift-independent indicator. Estimates are based on either new observations, or old observations newly re-analyzed. To find eligible distance estimates in the astronomical literature, we at first relied solely on keyword-based searches of the joint Smithsonian Astrophysical Observatory (SAO) and NASA Astrophysics Data System (ADS), described by Kurtz et al. (2000). ADS offers up-to-date access to the astronomical literature published in major astronomical and scientific journals, including *Astronomy & Astrophysics, Astronomical Journal, Astrophysical Journal, Monthly Notices of the Royal Astronomical Society, Publications of the Astronomical Society of the Pacific,* as well as *Nature*, and *Science*, and most other journals. As with other NED data, distance estimates are linked to the original sources using the ADS bibliographic reference code. These 19-digit identifiers are used globally to identify individual astronomical references, and were developed in collaboration between the Centre de Données astronomiques de Strasbourg (CDS) and NED (Schmitz et al. 1995).

Our first compilation provided some 3,000 mostly primary distance estimates for about 1,000 galaxies published in over 300 references spanning 1990 to 2005, inclusive. Since then, NED-D has grown and evolved in concert with the publication of extragalactic distances in the literature. For example, numerous large compilations providing distance estimates for thousands of galaxies based on secondary indicators have been and continue to be added. Growth in terms of the number of individual distance estimates, and galaxies with such estimates, as well as references cited is evident in the log of major updates to NED-D over the last decade, presented in Table 1.

In addition to ADS searches of papers with the keywords "galaxy + distance" published from 1980 to present, daily search of new astronomical papers published on the arXiv.org e-print service has been conducted since 2005. NED-D, therefore, is a comprehensive tabulation of the primary distance estimates published from 2005 to 2015 inclusive, and we will strive to keep it comprehensive as new data are published. For the period prior to 2005 (1980 to 2004), NED-D is as comprehensive as possible. Some distance estimates are published in papers that do not include the keywords "galaxies - distances", as said. In attempting to obtain these, we have had to expand the search methods further. The most conventional methods involve following the "trails" of references and authors in all articles that give previously published distance estimates. Although time consuming, this effort will remain important until replaced with automated methods that are sufficiently advanced and proven to be more effective. (See Section 6.)

NED-D also maintains comprehensive coverage of redshift-independent distances for high-redshift objects, including but not limited to, gamma-ray bursts at redshifts up to and beyond $z = 6$, and Type Ia supernovae, currently reaching $z \sim 2$. In 2012, NED-D was expanded to include a list of 290 historical distance estimates for 60 galaxies, namely those published by Hubble, Lundmark, and de Sitter, among thirty authors who published distance estimates from 1840 to 1930. Re-publication of the historical data was useful in confronting numerous articles published in the prior two years that cast doubt on Hubble's legacy in discovery of the expanding universe (e.g., van den Bergh 2011, Livio 2011). The data played a key role in confirming that while Lemaitre, Lundmark, and others deserve priority for discovering observational evidence for universal expansion, Hubble alone discovered observational proof (Steer 2011, 2012).

At the time of writing, approximately 10% of the distance estimates in NED-D were published in non-peer-reviewed sources such as conference proceedings. Those non-peer-reviewed estimates have been included because they come from established authors, have impact on thorough extragalactic research, and are un-obtainable in peer-reviewed publications.

For the LMC, the Milky Way satellite galaxy often referenced as the anchor or zero-point for the extragalactic distance scale, NED-D currently provides more than 1,300 distance estimates based on 30 primary indicators. Hundreds of primary estimates are likewise available for other Local Group galaxies, including the SMC (n > 600), and Messier 33 (n > 150). Many of these estimates provide distances to individual stars and other components within Local Group galaxies that probe the depth and 3D structure of these galaxies.

For 39 galaxies with Messier designations, more than 2,000 distance estimates based on primary and secondary indicators are currently available or an average of 55 estimates per galaxy. For 35 Messier galaxies, over 1,600 estimates are available based purely on primary indicators for an average of 47 primary estimates per galaxy. Among 6,067 galaxies in the New General Catalogue (NGC), around 60% (n = 3,625) have distances based on primary and secondary indicators, totaling 30,000 estimates or an average of 8 estimates per galaxy, and more than 13% (n = 798) have distances based purely on primary indicators, with 6,800 primary estimates for around 9 primary estimates per galaxy.

The most abundantly applied primary distance indicator is Type Ia supernovae, with estimates available for more than 3,000 galaxies. Major Type Ia supernovae distance compilations are included, including the Gold and Silver (Riess et al. 2004, n = 186), Constitution (Hicken et al. 2009, and references therein, n = 496), Union 2 (Amanullah et al. 2010, n = 687), and Sloan Digital Sky Survey (SDSS) II Supernova Survey (Sako et al. 2014, n ~ 500 confirmed and n > 2,000 including candidates).

The most abundantly applied secondary distance indicator is the Tully-Fisher relation, available for ~12,000 galaxies. Major Tully-Fisher distances catalogs are provided, including the Nearby Galaxies Catalog (NGC, Tully 1988, n = 2,371), the Mark III Catalog of Galaxy Peculiar Velocities (Mk III, Willick et al. 1997, n = 2,979), the Revised Flat Galaxies Catalog (RFGC, see Karachentsev et al. 2000, n = 1,327, and Parnovsky et al. 2010, n = 1,623), the Two-Micron All-Sky Survey (2MASS) Flat Galaxy Catalog (2MFGC, Karachentsev et al. (2006, 2011), n = 2,724), the Kinematics of the Local Universe survey (KLUN, Theureau et al. 2007, n = 4,149), and the Spiral Field I-band Plus survey (SFI++, Springob et al. 2009, and references therein, n = 4,857). Additional Tully-Fisher distances among the many thousands in the Extragalactic Distances Database mentioned earlier but not published in other references are also included (EDD, Tully et al. (2009, 2013), n = 1,023).

3. HOW NED-D IS FORMATTED AND MADE ACCESSIBLE

The complete compilation of redshift-independent distances is available for download on NED-D's main page[1]. The format is comma-separated value (CSV), appropriate for loading into most database or analysis tools. The formatting of NED-D is described in Table 2. In brief, for each distance estimate we provide the name of the galaxy the estimate applies to, using the NED Preferred Object Name for easy interaction with the main NED database. Distance estimates are given as a distance modulus (m-M), followed by the error (err) when available, the linear (proper motion) distance in megaparsecs (D(Mpc)), and the distance indicator used. Also given is the bibliographic reference code. Distance estimates in the full compilation not yet integrated into NED database queries are indicated in the first column, and noted as due to either references (R) or objects (N) not yet folded into the main database. Object names from the literature are listed for the latter. Note the full tabulation generally includes around 15% more data than is available for query via the main NED database. Most of the additional data involves non-peer-reviewed references mentioned which are excluded from the main database, while some involves peer-reviewed references not yet incorporated in the main database.

---

[1] http://ned.ipac.caltech.edu/Library/Distances/

NED-D is unique in allowing researchers to place estimates onto a level playing field, by accounting for differences among certain indicators and for differences in distance scale. Many Type Ia and some Type II supernovae estimates for example, are based not on the linear or proper motion distance modulus, m-M = 5logD-5, with D in parsecs (pc), but rather on the luminosity distance modulus, m-M = 5logD-5/(1+z), e.g., Lang (1980), Hogg (1999). The first ancillary column of NED-D is therefore used to indicate supernovae-based estimates by providing the supernova ID. The second ancillary column indicates where estimates based on supernovae and other indicators including but not limited to gamma ray bursts, are based on luminosity distance moduli, by providing the target "redshift (z)". Note for ease of use, researchers can simply refer to the linear distances given in Mpc, where the difference between proper motion- and luminosity-based distance moduli is already accounted for.

For researchers requiring maximum precision, the third ancillary column provides the "Hubble constant" assumed and indicates cases where this differs from the default value of $H_0 = 70$ km s$^{-1}$ Mpc$^{-1}$ based on the NASA HST Key Project final result of $H_0 = 72 \pm 8$ km s$^{-1}$ Mpc$^{-1}$ (Freedman et al. 2001), rounded to the nearest 5 km s$^{-1}$ Mpc$^{-1}$. For example, Type Ia supernovae estimates provided by the Sandage & Tammann group (Reindl et al. 2005), the Riess et al. group (Riess et al. 2004), and the Perlmuter et al. group (Amanullah et al. 2010), have hitherto been tied to values of $H_0 = 60$, 65, and 70 km s$^{-1}$ Mpc$^{-1}$, respectively. It is important to note that all extragalactic distance indicators provide only relative distances. In the case of some high-redshift indicators, such as Type Ia supernovae, different Hubble constants are used to provide different zero points to convert relative distances into absolute distances. Other extragalactic distance indicator zero points of choice, including the distance to the LMC, are too nearby to be directly tied to distances based on Type Ia supernovae, because the nearest examples after three decades all lie beyond 3.5 Mpc.

To place estimates on a uniform scale, they must first be sorted by the Hubble constant assumed, and estimates affected by differences from the default value must be standardized. Once placed on a uniform basis with the default value, mean distance(s) can be obtained. Further, once standardized to the default value, estimates can be adjusted *en masse* to obtain mean distance(s) based on any value of the Hubble constant.

The fourth ancillary column gives the "Adopted LMC modulus," and where applicable, indicates where a distance scale zero point has been assumed that is different from that selected by the NASA HST Key Project, i.e., an LMC distance modulus of m-M = 18.50 mag and a linear distance of 50.1 kpc (Freedman et al. 2001). To place all estimates on a uniform scale, the estimates must first be sorted by the "Adopted LMC modulus", and estimates affected by differences from the default value must be standardized. As with the Hubble constant, once standardized to the default value, mean distance(s) can be obtained and then based on any value of the LMC zero point.

One further ancillary column in the full compilation provides the number of years since 1980 for publication date, for date weighting of estimates. Newer estimates are generally more accurate, benefiting from both improved techniques and experience. Weighting by published errors, a common practice, is doable by referring to the distance modulus errors where available, but should be done with caution. Published errors are known to be severely heterogeneous (Tammann et al. 1991, Rubin et al. 2015).

Distance estimates in the full compilation are given for each galaxy in order of distance indicator used and within each indicator in increasing distance. Individual galaxies are given in order of Right Ascension. An Estimate Number column and a Galaxy Number column are also provided. Distances are based on 75 different indicators currently, with 51 primary indicators, including 41 standard candle and 10 standard ruler indicators, and 24 secondary indicators. Available indicators, and the number of estimates and galaxies each applies to, are shown in Table 3. Number of references, authors, and citations for each indicator, as well as the minimum, mean, and maximum distances obtained, and their mean statistical error are also shown. Descriptions of each indicator in use at time of writing, including explanatory examples of applications to galaxies in the database, are given in the Appendix. Updated descriptions, including of new indicators as they become available are accessible online, by following the links in the online table of indicators on NED-D's main page.

In addition to original distance estimates, certain estimates have been repeated because they provide distances to individual stars and other components within galaxies. These can include Cepheids within the LMC, globular clusters within Messier 31, or supernovae within galaxies including four Type Ia supernovae within NGC 1316, the central galaxy of the Fornax cluster of galaxies. All examples of repeated distances for objects within galaxies follow the galaxies they apply to. They are distinguished from original estimates in the galaxy ID column and in the notes column respectively, by showing the individual objects in the first and the galaxies they are within in the latter, e.g., the Type II supernova object SN 1987A in the galaxy LMC. They are further distinguished by the number "999999" in the Estimate Number column. Distances to individual objects within galaxies are useful for studying the depth and 3D structure of the galaxies they reside in. Such distances are available for Local Group galaxies, and most abundantly available for the LMC and SMC. These include individual distances to 136 Cepheids (Pejcha & Kochanek 2012), 56 planetary nebulae (Ortiz 2013), and 50 RR Lyrae stars (Borissova et al. 2009) in the LMC and 53 Cepheids (Pejcha & Kochanek 2012) and 32 Eclipsing Binaries (North et al. 2010) in the SMC. Recently published distances for 8,876

Cepheids, including 4,222 in the LMC and 4,654 in the SMC (Jacyszyn-Dobrzeniecka et al. 2016) are in the process of being included. Users of the full compilation interested only in unique original distance estimates can easily retrieve those for individual galaxies by referring to the Estimate Number column, and avoiding estimates marked "999999," and for galaxies *en masse* by sorting the full compilation by Estimate Number column and eliminating repeated estimates, which labeled "999999" all follow the highest estimate number.

Redshift-independent distances for individual galaxies can be obtained by using the "Redshift-Independent Distances" service available on NED's main interface page[2]. Enter the object name and a mean distance for any galaxy with such distances is reported, along with a statistical summary and a summary table of the available estimates, based on estimates published in peer-reviewed references currently included in the main database.

For galaxy Messier 31, for example, this option currently returns more than 300 estimates based on over a dozen primary indicators. This option also provides summary statistics for the distribution of all available distance estimates, including the mean, median, minimum, maximum, and standard deviation. An example is shown in Figure 1.

Mean distances and errors can also be found, where available, in each individual galaxy's data summary page. Mean distances for large numbers of galaxies can be retrieved via the Build Data Table service on NED's main interface. Either the names or coordinates of galaxies may be entered as input. Large numbers of galaxies with redshift-independent distances can also be accessed by searching NED for objects via Classifications, Types, and Attributes, and selecting objects with distances based on specific indicators (e.g., Tully-Fisher). In deriving quantities such as metric sizes (e.g., in kpc) and absolute magnitudes or luminosities, NED uses a redshift-independent distance where available rather than a redshift-based distance. Users of NED mean distances are cautioned however that currently, our summary statistics are based on original values as published. No homogenization or corrections have been applied.

The purpose of the all-inclusive approach of NED-D is to allow astronomers to compare, analyze and filter all available data in any way they wish. The importance of this is best illustrated by the megamaser-based distance estimate to galaxy Messier 106 (NGC 4258), considered the most accurate distance estimate to date beyond the Local Group, ± 4%. Eschewing older data as obsolete, some researchers will consider only the most recent megamaser-based estimate, $D = 7.6 \pm 0.3$ Mpc (Humphreys et al. 2013). However, when all data are considered, it is evident that megamaser-based estimates for Messier 106 have undergone a systematic increase, from 6.4 Mpc (Miyoshi et al. 1995) to 7.2 Mpc (Herrnstein et al. 1999), and from 7.3 Mpc (Riess et al. 2012, based on private communication with Humphreys) to the current value of 7.6 Mpc by Humphreys et al. (2013). Only by having all available data can researchers make informed checks on estimates of the distances to particular galaxies, and estimates of the Hubble constant based on distances to numerous galaxies.

---

[2] http://ned.ipac.caltech.edu

## 4. DISTANCES DATA GROWTH, APPLICABILTY, AND RESEARCH ACTIVITY

In this section, we summarize the growth in redshift-independent distances since 1980, the applicability of the data available, and the research activity producing the data.

*4.1 Growth in data*

The number of distance estimates published based on primary indicators is doubling approximately every four years, as shown in Figure 2. Growth by a factor of 100x in primary estimates over the last thirty years means that currently, more than 20,000 such estimates are available compared to ~200 published from 1980 to 1985. Indeed, growth in primary estimates is occurring at a remarkably constant rate. Estimates available have doubled every four years for the last three decades. Growth in secondary estimates has been broadly parallel to, but less steady than in primary estimates. Updates to NED-D are made regularly, and the cadence has increased recently from 4 to 6 releases per year. In general, the full tabulation appearing online is current to within the last 6 months.

Growth in data is attributable to corresponding growth in extragalactic distance scale research activity. Growth in the number of references published with primary distances, the number of estimates published per reference, and the number of authors per reference is shown in Figure 3. Specifically, growth in primary distances estimates of 100x over the past three decades is due in small part to growth in the number of references published per year (5x), and in large part due to growth in the number of estimates per reference (20x). For the period 2012 to 2014 inclusive, primary distance estimates per reference have averaged 60. That is as many primary distances per reference today as were published in the NASA HST Key Project final report, which gave 62 Cepheids-based distances for 31 galaxies (Freedman et al. 2001).

*4.2 Applicability*

A graphical presentation that facilitates comparison of the distance indicators is shown in Figure 4. Each indicator applies to a range of distances, as indicated by boxes showing the 25th percentile (left side), 50th percentile (median, inner solid line), and 75th percentile (right side), along with "whiskers" spanning the minimum and maximum values. Figure 4 reveals some indicators with distance estimates as far as, or even beyond, the ~14 Gpc radius of the observable universe based on the standard cosmological model (e.g., Lineweaver & Davis 2005, noting that radius is ~3x greater than the simplistic Hubble radius of 4.3 Gpc based on $r_H = c/H_0$ and assuming $H_0 = 70$ km s$^{-1}$ Mpc$^{-1}$). Some extremely large distances involve indicators with very high errors, including gamma ray bursts (GRB), gravitational lenses (G Lens), and HII Luminosity Function (HII LF), which can be over-estimated (and in other cases under-estimated) by factors of 2.5 or more. Some are based on Type Ia supernovae, and evidently represent statistical outliers or tentative identifications (candidates rather than confirmed) (SNIa, and SNIa SDSS). Rather than censoring such outliers, they are included in the database with their corresponding errors to accurately reflect what has been collated from the peer-reviewed literature to date. Clearly much work remains to decrease error in the techniques, and to adjust expectations regarding realistic ranges of applicability of these specific distance indicators.

Indicators applicable to cosmological distances, such as Type Ia supernovae and GRBs, are severely limited in the minimum distance they apply to. The nearest Type Ia supernova in three decades, SN 2014J, occurred at 3.5 Mpc. The primary concern of extragalactic distance scale research is to use nearer redshift-independent indicators to calibrate more distant, cosmological indicators, in order to fix the extragalactic distance scale more precisely, and to obtain cosmological parameters with more precision.

*4.3 Research activity*

Specific distance indicators have a perceived quality relative to each other that is subject to interpretation. The activity researchers have devoted to each indicator however, is a matter of record. Research activity in terms of the number of references and number of authors devoted to each indicator, and the number of citations received by each since 1980, is shown in Figure 5. When the indicators are shown in increasing order of the number of references that have applied each indicator to date, the research activity in terms of numbers of authors contributing, estimates published, and citations received are clearly correlated. Specific indicators appearing toward the top of Figure 5 are in general therefore, more "tried and true", compared to indicators appearing toward the bottom. At the time of writing, the top 5 most applied standard candle indicators by number of references are: Cepheids, tip of the red giant branch, RR Lyrae, color-magnitude diagram, and Type Ia supernovae. The top 3 most applied standard rulers are: masers, eclipsing binary, and Type II supernovae (optical). The top 3 most applied secondary indicators are: Tully-Fisher, fundamental plane, and diameter-rotational velocity (D_n-sigma). Note that some indicators have little research activity but many citations, because although rarely applied, they were published for comparison with more widely used indicators in articles that garner more citations.

5. USAGE AND FUTURE PROSPECTS

In this section we provide a brief summary of NED-D usage in the literature to date, and discuss future prospects in terms of improving accessibility, usability, and impact of the data.

*5.1 Usage*

Usage of NED-D has been cited in nearly 250 astronomical articles as of Jan. 1, 2016. References in the year 2012 totaled 46, up from only 2 or 3 references per year in NED-D's first two years. Citations in 2014, the most recent full year considered, are close to 60. NED-D has been cited in more than 200 papers published in the top 7 astronomical and science publications, as measured by impact factors. Examples include:

McCommas et al. (2009), Freedman & Madore (2010), Burns et al. (2011), Freedman et al. (2011), Smith et al. (2012), Hunter et al. (2012), Jarrett et al. (2013), Pietrzynski et al. (2013), Petty et al. (2014), and White et al. (2015).

*5.2 Future Prospects*

Two core activities for NED-D, as for NED in general, involve improving the content and search capabilities, and updating to maintain the most complete data possible. In addition to keeping pace with new data appearing in the literature, NED-D completeness will improve by incorporating ~9,000 secondary distances from 50 references, which is in our near-term work plan. The challenge is that most of these data sources involve large lists and tables of galaxies that are not currently machine-readable because they are found in older, non-digitized articles published prior to 2000.

The percentage of references being published with eligible distance estimates but without keywords "galaxies - distances" in their abstracts is a concern, and worth attempting to correct. To do so, we are planning to add an update regarding publication of redshift-independent distances to our recently published Best Practices for Data Publication (Schmitz et al. 2014)[3]. Researchers publishing distance estimates for Type Ia supernovae, gamma-ray bursts, and other extragalactic objects are encouraged to include the keywords "galaxies - distances", as well as the object descriptors in their abstracts, so that NED and others interested in such estimates are able to easily locate them.

Greater inclusion, rather than exclusion of relevant data is important, because of the relative rarity and high value of redshift-independent distances data. Whether to include data in the main NED database from non peer-reviewed references available in NED-D, however, remains an open issue. Therefore, we invite user input on this issue[4]. In addition, the current practice of providing mean NED distances based on estimates as published, uncorrected for assumed distance scale, will be upgraded to include mean estimates accounting for author differences in scale.

The NED team is in the process of developing a number of new interactive data visualizations to facilitate understanding the database content, and to simplify new types of database queries. For example, the figures in this article are static snapshots of interactive visualizations (allowing zooming, panning, display of data attributes while hovering over markers, etc.) that are being configured on the NED website and will be updated as new content are added to the database.

Further plans include enhancing the NED Galaxy Environment service[5]. Introduced by NED in 2013, this feature allows researchers to quickly ascertain and graphically display the 3D neighbors of galaxies with available redshifts. A future version will include redshift-independent distances. We are also exploring techniques for generating interactive 3D maps of galaxy distributions using distances derived both from redshifts and redshift-independent indicators, for example using the World Wide Telescope (see Roberts & Fay 2014).

---

[3] http://ned.ipac.caltech.edu/docs/BPDP/NED_BPDP.pdf
[4] Please use the Contact Us or Comment option on NED's main page: http://ned.ipac.caltech.edu/forms/comment.html
[5] http://ned.ipac.caltech.edu/forms/denv.html

A related work in progress at NED involves identifying galaxy neighbors by their hierarchy, by recording and reporting on which galaxies have been identified in the literature as being members of pairs, groups, clusters, and superclusters. For NED-D, hierarchy information could be used as a force multiplier, multiplying by three or more times the number of galaxies with effective redshift-independent distances, albeit with cautionary flags to distinguish hierarchy-based inferred distances from original distance estimates.

Our ultimate plan to ensure the most complete coverage of data relevant to NED-D, as for other data types in NED over all, is to apply, test, and put into operation text data mining algorithms to locate, classify, tag, and simplify extraction of relevant data. Simply put, effective application of modern machine learning algorithms may be the only practical way to keep the database as comprehensive as possible amidst the rapid growth of data published annually in the literature. Initial steps in this area have begun.

## 6. SUMMARY

NED-D is designed to meet the need for an up-to-date, easy-to-use and comprehensive compilation of redshift-independent extragalactic distances. NED-D is being maintained as part of NED core activities to support scientists, space missions, and ground-based observatories in planning, interpretation, and publication of research on galaxies, extragalactic distances, and cosmology. Updated versions are provided on a regular basis.

As of January 2016, more than 100,000 separate distance estimates based on primary and secondary distance indicators are available for over 28,000 individual galaxies, and compiled from over 2,000 references. A decade old, growing rapidly, and based on keyword searches of the ADS, daily search of arXiv.org, and other search methods, NED-D offers a valuable reference to the redshift-independent extragalactic distance estimates published in the astronomical literature from 1980 to present.

Growth in the number of distance estimates published based on primary indicators appears close to constant, doubling approximately every four years. Over three decades, from 1985 to 2015, growth by a factor of 100 in primary-based estimates has been driven by growth in research activity. There are five times more references per year, 20 times more estimates per reference, and five times more authors per reference providing primary-based distance estimates today than there were thirty years ago.

The top 5 most applied standard candle indicators by number of references are: Cepheids, tip of the red giant branch, RR Lyrae, color-magnitude diagram, and Type Ia supernovae.

The top 3 most applied standard rulers are: masers, eclipsing binary, and Type II supernovae (optical). The top 3 most applied secondary indicators are: Tully-Fisher, fundamental plane, and diameter-rotational velocity (D_n-sigma).

NED-D is having a significant impact on assisting extragalactic research, as demonstrated by citations in nearly 250 astronomical articles as of 2016 January 1. This includes more than 200 articles published in the top 7 astronomical and science publications.

We encourage authors to include the keywords "galaxies - distances" in the abstracts of articles offering new extragalactic distance estimates, both for researchers interested in such distances and to smooth the process of keeping NED-D as complete as possible.

Future prospects include the use of redshift-independent distances (and other data in NED) in interactive visualizations of the database; queries and visualizations of galaxy environments and large scale structure in the universe; in the application of machine learning algorithms to locate, classify, and tag relevant measurements as they appear in the literature to keep the content as complete and current as possible; and in facilitating comparison of different distance indicators for more precise calibration of the extragalactic distance scale.

The authors are grateful to the many authors who publish redshift-independent extragalactic distance estimates. In particular we would like to thank Edward Baron, Jonathan Bird, Massimiliano Bonamente, Dmitry Bizyaev, John Blakeslee, Jean Brodie, Heather Campbell, Chris Corbally, Igor Drozdovsky, Wendy Freedman, Mohan Ganeshalingam, Gretchen Harris, William Harris, Raoul Haschke, Martha Haynes, Robert Hurt, George Jacoby, Igor Karachentsev, David Lagattuta, Tod Lauer, Mario Livio, Lucas Macri, Daniel Majaess, Dmitry Makarov, Karen Masters, Kristen McQuinn, Fulvio Melia, Jeremy Mould, Robert Quimby, Armin Rest, Adam Riess, Luca Rizzi, David Russell, Christoph Saulder, Riccardo Scarpa, Bradley Schaefer, David Schlegel, Daniel Scolnic, Chris Springob, Gilles Theureau, Brent Tully, Alan Whiting, and Henrique Xavier for many helpful comments over the years. This research has made extensive use of the SAO/NASA Astrophysics Data System Bibliographic services. This work has also made extensive use of, and is funded by the NASA/IPAC Extragalactic Database (NED), which is operated by the Jet Propulsion Laboratory, California Institute of Technology, under contract with the National Aeronautics and Space Administration. Additional generous support to I.S. from the Carnegie Institution of Canada is also gratefully appreciated.

APPENDIX

DESCRIPTIONS AND EXAMPLES OF DISTANCE INDICATORS IN NED-D

For further information and updates to this material, see
http://ned.ipac.caltech.edu/Library/Distances/distintro.html

Descriptions of distance indicators that follow are brief. The references were chosen randomly from uses in NED-D, and are provided only as illustrative examples. For in-depth reviews of specific indicators or to obtain references giving the original, first uses of indicators, follow the references given and the references therein. For in-depth reviews on primary indicators see Ferrarese et al. (2000), Freedman & Madore (2010), de Grijs et al. (2014) and de Grijs & Bono (2015, 2014, and references therein), and for secondary indicators see Tully et al. (2009, 2013, 2016, and references therein).

Descriptions of standard candle indicators are given in Section A.1, followed by standard ruler indicators in Section A.2, and secondary indicators in Section A.3. Additional information on applying Cepheids in particular, and applicable to standard candle-based indicators in general, is given in Section A.4 . Brief descriptions of luminosity relations, apparent versus reddening-corrected distance, and corrections related to age or metallicity, as well as others are provided.

Researchers are cautioned that at least three indicators have considerable overlap with others. Asymptotic Giant Branch (AGB) stars are a particular type of brightest stars indicator. The Subdwarf Fitting indicator makes use of the CMD indicator, but is applied specifically to globular clusters. The Dwarf Elliptical indicator makes use of the better-known fundamental plane relation for elliptical galaxies, but is applied specifically to dwarf elliptical galaxies. The indicators mentioned are considered distinct empirically, because they pertain to different stellar populations. They are treated as distinct in the references provided for the indicators, and in the literature in general. Further, distinguishing indicators based on the stellar populations targeted is in keeping with recognition of the TRGB, Horizontal Branch, and Red Clump indicators as distinct indicators, though all are related to the CMD indicator.

## A.1. Standard Candles

AGN Time lag: based on the time lag between variations in magnitude observed at short wavelengths compared to those observed at longer wavelengths in Active Galactic Nuclei (AGN). For example, using a quantitative physical model that relates the time lag to the absolute luminosity of an AGN, Yoshii et al. (2014) obtain a distance to the AGN host galaxy MRK 0335 of 146 Mpc.

Asymptotic Giant Branch Stars (AGB): based on the maximum absolute visual magnitude for these stars of $M_V = -2.8$ (Davidge & Pritchet 1990). Thus, the brightest AGB Stars in the galaxy NGC 0253, with a maximum apparent visual magnitude of $m_V = 24.0$, have a distance modulus of $(m-M)_V = 26.8$, for a distance of 2.3 Mpc.

B-type Stars (B Stars): based on the relation between absolute-magnitude and beta-index in these stars, where beta-index measures the strength of the star's emission at the wavelength of Hydrogen Balmer or H-Beta emission. Applied to the LMC by Shobbrook & Visvanathan (1987), to obtain a distance modulus of (m-M) = 18.30, for a distance of 46 kpc, with a statistical error of 0.20 mag or 4 kpc (10%).

BL Lac Object Luminosity (BL Lac Luminosity): based on the mean absolute magnitude of the giant elliptical host galaxies of these Active Galactic Nuclei (AGN). Applied to BL Lacertae host galaxy MS 0122.1+0903 by Sbarufatti et al. (2005), to obtain a distance of 1,530 Mpc.

Black Hole: based on super-Eddington accreting massive black holes, as found the host galaxies of certain Active Galactic Nuclei (AGN) at high redshift, and a unique relationship between their bolometric luminosity and central black hole mass. Based on a method to estimate black hole masses (Wang et al. 2014), the black hole mass-luminosity relation is used to estimate the distance to 16 AGN host galaxies, including for example galaxy MRK 0335, to obtain a distance of 85.9 Mpc, with a statistical error of 26.3 Mpc (31%).

Blue Supergiant: based on the absolute magnitude and the equivalent widths of the Hydrogen Balmer lines of these stars. Applied to the SMC by Bresolin (2003) to obtain a distance modulus of (m-M) = 19.00, for a distance of 64 kpc, with a statistical error of 0.50 mag or 16 kpc (25%).

Brightest Stars: based on the mean absolute visual magnitude for red supergiant stars, $M_V$ = -8.0, Davidge, Le Fevre, & Clark (1991) present an application to NGC 0253 where red supergiant stars have apparent visual magnitude $m_V$ = 19.0, leading to a distance modulus of $(m-M)_V$ = 27.0, for a distance of 2.5 Mpc.

Carbon Stars: based on the mean absolute near-infrared magnitude of these stars $M_I$ = -4.75 (Pritchet et al. 1987). Thus, Carbon Stars in galaxy NGC 0055 with a maximum apparent infrared magnitude $m_I$ = 21.02, including a correction of -0.11 mag for reddening, have a distance modulus of $(m-M)_I$ = 25.66, for a distance of 1.34 Mpc, with a statistical error of 0.13 mag or 0.08 Mpc (6%).

Cepheids: based on the mean luminosity of Cepheid variable stars, which depends on their pulsation period, P. For example, a Cepheid with a period of P = 54.4 days has an absolute mean visual magnitude of $M_V$ = -6.25, based on the period-luminosity (PL) relation adopted by the HST Key Project on the Extragalactic Distance Scale (Freedman et al. 2001). Thus, a Cepheid with a period of P = 54.4 days in the galaxy NGC 1637 (Leonard et al. 2003) with an apparent mean visual magnitude $m_V$ = 24.19, has an apparent visual distance modulus of $(m-M)_V$ = 30.44, for a distance of 12.2 Mpc. Averaging the apparent visual distance moduli for the eighteen Cepheids known in this galaxy (including corrections of 0.10 mag for reddening and metallicity) gives a corrected distance modulus of $(m-M)_V$ = 30.34, for a distance of 11.7 Mpc, with a statistical error of 0.07 mag or 0.4 Mpc (3.5%).

Color-Magnitude Diagrams (CMD): based on the absolute magnitude of a galaxy's various stellar populations, discernable in a color-magnitude diagram. Applied to the LMC by Andersen et al. (1984), to obtain a distance modulus of (m-M) = 18.40, for a distance of 47.9 kpc.

Delta Scuti: based on the mean absolute magnitude of these variable stars, which depends on their pulsation period. As with Cepheid and Mira variables, a period-luminosity (PL) relation gives their absolute magnitude. Applied to the LMC by McNamara et al. (2007), to obtain a distance modulus of (m-M) = 19.46, for a distance of 49 kpc, with a statistical error of 0.19 mag or 4.5 kpc (9%).

Flux-Weighted Gravity-Luminosity Relation (FGLR): based on the absolute bolometric magnitude of A-type supergiant stars, determined by the Flux-Weighted Gravity- Luminosity Relation (Kudritzki et al. 2008). Applied to galaxy Messier 31, to obtain a distance of 0.783 Mpc.

Gamma-Ray Burst (GRB): based on six correlations of observed properties of GRBs with their luminosities or collimation-corrected energies. A Bayesian fitting procedure then leads to the best combination of these correlations for a given data set and cosmological model. Applied to GRB

021004 by Cardone et al. (2009), to obtain a luminosity distance modulus of (m-M) = 46.60 for a luminosity distance of 20,900 Mpc. With the GRB's redshift of z = 2.3, this leads to a linear distance of 6,330 Mpc, with a statistical error of 0.48 mag or 1,570 Mpc (25%).

Globular Cluster Luminosity Function (GCLF): based on an absolute visual magnitude of $M_V$ = -7.6, which is the location of the peak in the luminosity function of old, blue, low-metallicity globular clusters (Larsen et al. 2001). So, for example, the galaxy NGC 0524 with an apparent visual magnitude $m_V$ = 24.36 for the peak in the luminosity function of its globular clusters, has a distance modulus of $(m-M)_V$ = 31.99, for a distance of 25 Mpc, with a statistical error of 0.14 mag or 1.8 Mpc (7%).

Globular Cluster Surface Brightness Fluctuations (GC SBF): based on the fluctuations in surface brightness arising from the mottling of the otherwise smooth light of the cluster due to individual stars (Ajhar et al. 1996). Thus, the implied apparent magnitude of the stars leading to these fluctuations gives the distance modulus in magnitudes. Applied to galaxy Messier 31, to obtain a distance modulus of (m-M) = 24.56, for a distance of 0.817 Mpc, with a statistical error of 0.12 mag or 0.046 Mpc (6%).

HII Luminosity Function (HII LF): based on a relation between velocity dispersion, metallicity, and the luminosity of the H-beta line in HII regions and HII galaxies (e.g., Siegel et al. 2005, and references therein). Applied to high-redshift galaxy CDFa C01, to obtain a luminosity distance modulus of (m-M) = 45.77, for a luminosity distance of 14,260 Mpc. With a redshift for the galaxy of z = 3.11, this leads to a linear distance of 3,470 Mpc, with a statistical error of 1.58 mag or 3,710 Mpc (93%).

Horizontal Branch: based on the absolute visual magnitude of horizontal branch stars, which is close to $M_V$ = +0.50, but depends on metallicity (Da Costa et al. 2002). Thus, horizontal branch stars in the galaxy Andromeda III with an apparent visual magnitude $m_V$ = 25.06, including a reddening correction of -0.18 mag, have a distance modulus of $(m-M)_V$ = 24.38, for a distance of 750 kpc, with a statistical error of 0.06 mag or 20 kpc (3%).

M Stars luminosity (M Stars): based on the relationship between absolute magnitude and temperature-independent spectral index for normal M Stars. Applied to the LMC by Schmidt-Kaler & Oestreicher (1998), to obtain a distance modulus of (m-M) = 18.34, for a distance of 46.6 kpc, with a statistical error of 0.09 mag or 2.0 kpc (4%).

Miras: based on the mean absolute magnitude of Mira variable stars, which depends on their pulsation period. As with Cepheid variables, a period-luminosity (PL) relation gives their absolute magnitude. Applied to the LMC by Feast et al. (2002), to obtain a distance modulus of (m-M) = 18.60, for a distance of 52.5 kpc, with a statistical error of 0.10 mag or 2.5 kpc (5%).

Novae: based on the maximum absolute visual magnitude reached by these explosions, which is $M_V$ = -8.77 (Ferrarese et al. 1996). So, a nova in galaxy Messier 100 with a maximum apparent visual magnitude of $m_V$ = 22.27, has a distance modulus of $(m-M)_V$ = 31.0, for a distance of 15.8 Mpc, with a statistical error of 0.3 mag or 2.4 Mpc (15%).

O- and B-type Supergiants (OB Stars): based on the relationship between spectral type, luminosity class, and absolute magnitude for these stars. Applied to 30 Doradus in the LMC by Walborn & Blades (1997), to obtain a distance of 53 kpc.

Planetary Nebula Luminosity Function (PNLF): based on the maximum absolute visual magnitude for planetary nebulae of $M_V$ = -4.48 (Ciardullo et al. 2002). So, planetary nebulae in the galaxy NGC 2403 with a maximum apparent visual magnitude of $m_V$ = 23.17 have a distance modulus of $(m-M)_V$ = 27.65, for a distance of 3.4 Mpc, with a statistical error of 0.17 mag or 0.29 Mpc (8.5%).

Post-Asymptotic Giant Branch Stars (PAGB Stars): based on the maximum absolute visual magnitude for these stars of $M_V$ = -3.3 (Bond & Alves 2001). Thus, PAGB Stars in Messier 31 with a maximum apparent visual magnitude of $m_V$ = 20.88 have a distance modulus of $(m-M)_V$ = 24.2, for a distance of 690 kpc, with a statistical error of 0.06 mag or 20 kpc (3%).

Quasar spectrum: based on the observed apparent spectrum of a quasar, compared with the absolute spectrum of comparable quasars as determined based on Hubble Space Telescope spectra taken of 101 quasars. Applied to 11 quasars by de Bruijne et al. (2002), including quasar [HB89] 0000-263, to obtain a distance of 3.97 Gpc.

RR Lyrae Stars: based on the mean absolute visual magnitude of these variable stars, which depends on metallicity: $M_V$ = F/H x 0.17 + 0.82 mag (Pritzl et al. 2005). So, RR Lyrae stars with metallicity F/H = -1.88 in the galaxy Andromeda III have an apparent mean visual magnitude of $m_V$ = 24.84, including a 0.17 mag correction for reddening. Thus, they have a distance modulus of $(m-M)_V$ = 24.34, for a distance of 740 kpc, with a statistical error of 0.06 mag or 22 kpc (3.0%).

Red Clump: based on the maximum absolute infrared magnitude for Red Clump stars of $M_I$ = -0.67 (Dolphin et al. 2003). So, red clump stars in the galaxy Sextans A with a maximum apparent infrared magnitude of $m_I$ = 24.84, including a 0.07 mag correction for reddening, have a distance modulus of $(m-M)_I$ = 25.51, for a distance of 1.26 Mpc, with a statistical error of 0.15 mag or 0.09 Mpc (7.5%).

Red Supergiant Variables (RSV Stars): based on the mean absolute magnitude of these variable stars, which depends on their pulsation period (Jurcevic 1998). As with Cepheid and Mira variables, a period-luminosity (PL) relation gives their absolute magnitude. Applied to galaxy NGC 2366, to obtain a distance modulus of (m-M) = 27.86, for a distance of 3.73 Mpc, with a statistical error of 0.20 mag or 0.36 Mpc (10%).

Red Variable Stars (RV Stars): based on the mean absolute magnitude of RV Stars, which depends on their pulsation period (Kiss & Bedding 2004). As with Cepheid variables, a period-luminosity (PL) relation gives their absolute magnitude. Applied to the SMC to obtain a distance modulus of (m-M) = 18.94, for a distance of 61.4 kpc, with a statistical error of 0.05 mag or 1.4 kpc (2.3%)

S Doradus Stars: based on the mean absolute magnitude of these stars, which is derived based on their amplitude-luminosity relation. Applied to galaxy Messier 31 by Wolf (1989), to obtain a distance modulus of (m-M) = 24.40, for a distance of 0.759 Mpc.

SNIa SDSS: based on SNIa (Type Ia supernovae). It is distinguished from normal SNIa however, because it has been applied to candidate SNIa obtained in the SDSS Supernova Survey that have not yet been confirmed as bona fide SNIa (Sako et al., 2014). Applied to Type Ia supernova SDSS-II SN 13651, to obtain a luminosity distance modulus of (m-M) = 41.64 for a luminosity distance of 2,130 Mpc. With a redshift for the supernova of z = 0.25, this leads to a linear distance of 1,700 Mpc.

SX Phoenicis Stars: based on the mean absolute magnitude of these variable stars, which depends on their pulsation period. As with Cepheid and Mira variables, a period-luminosity (PL) relation gives their absolute magnitude (e.g., McNamara 1995). Applied to the Carina Dwarf Spheroidal galaxy, to obtain a distance modulus of (m-M) = 20.01, for a distance of 0.100 Mpc, with a statistical error of 0.05 mag or 0.002 Mpc (2.3%).

Short Gamma-Ray Bursts (SGRB): similar to but distinct from the GRB standard candle, because it employs only GRBs of short, less than 2 second duration (Rhoads 2010). SGRBs are conjectured to be a distinct subclass of GRBs, differing from the majority of normal or "long" GRBs, which have durations of greater than 2 seconds. Applied to SGRB GRB 070724A, to obtain a linear distance of 557 Mpc.

Statistical: based on the mean distance obtained from multiple distance estimates, based on at least several to as many as a dozen or more different standard candle indicators, although standard ruler indicators may also be included. For example, Freedman & Madore (2010) analyzed 180 estimates of the distance to the Large Magellanic Cloud, based on two dozen indicators not including Cepheids, to obtain a mean distance modulus of (m-M) = 18.44, for a distance of 48.8 kpc, with a statistical error of 0.18 mag or 4.2 kpc (9%).

Subdwarf Fitting: gives an improved calibration of the distances and ages of globular clusters. Applied to the LMC by Carretta et al. (2000), to obtain a distance modulus of (m-M) = 18.64, for a linear distance of 53.5 kpc, with a statistical error of 0.12 mag or 3.0 kpc (6%).

Sunyaev-Zeldovich Effect (SZ effect): based on the predicted Compton scattering between the photons of the cosmic microwave background radiation and electrons in galaxy clusters, and the observed scattering, giving an estimate of the distance. For galaxy cluster CL 0016+1609, Bonamente et al. (2006) obtain a linear distance of 1,300 Mpc, assuming an isothermal distribution.

Surface Brightness Fluctuations (SBF): based on the fluctuations in surface brightness arising from the mottling of the otherwise smooth light of the galaxy due to individual stars, primarily red giants with maximum absolute K-band magnitudes of $M_K$ = -5.6 (Jensen et al. 1998). So, the galaxy NGC 1399, for example, with brightest stars at an implied maximum apparent K-band magnitude $m_K$ = 25.98, has a distance modulus of $(m-M)_K$ = 31.59, for a distance of 20.8 Mpc, with a statistical error of 0.16 mag or 1.7 Mpc (8%).

Tip of the Red Giant Branch (TRGB): based on the maximum absolute infrared magnitude for TRGB Stars of $M_I$ = -4.1 (Sakai et al. 2000). So, the LMC, with a maximum apparent infrared magnitude for these stars of $m_I$ = 14.54, has a distance modulus of $(m-M)_I$ = 18.59, for a distance of 52 kpc, with a statistical error of 0.09 mag or 2 kpc (4.5%).

Type II Cepheids: based on the mean absolute magnitude of these variable stars, which depends on their pulsation period. As with normal Cepheids and Miras, a period-luminosity (PL) relation gives their absolute magnitude. Applied to galaxy NGC 4603 by Majaess et al. (2009), to obtain a distance modulus of (m-M) = 32.46, for a linear distance of 31.0 Mpc, with a statistical error of 0.44 mag or 7.0 Mpc (22%).

Type II Supernovae, Radio (SNII radio): based on the maximum absolute radio magnitude reached by these explosions, which is $5.5 \times 10^{23}$ ergs s$^{-1}$ Hz$^{-1}$ (Clocchiatti et al. 1995). So, the type-II SN 1993J in galaxy Messier 81 (NGC 3031), based on its maximum apparent radio magnitude, has a distance of 2.4 Mpc.

Type Ia Supernovae (SNIa): based on the maximum absolute blue magnitude reached by these explosions, which is $M_B$ = -19.3 (Astier et al. 2006). Thus, for example, SN 1990O (in the galaxy MCG +03-44-003) with a maximum apparent blue magnitude of $m_B$ = 16.20, has a luminosity distance modulus of $(m-M)_B$ = 35.54 (including a 0.03 mag correction for color and redshift), or a luminosity distance of 128 Mpc. With a redshift for the galaxy of z = 0.0307, this leads to a linear distance of 124 Mpc, with a statistical error of 0.09 mag or 6 Mpc (4.5%).

White Dwarfs: based on the absolute magnitudes of white dwarf stars, which depends on their age. Applied to the LMC by Carretta et al. (2000), to obtain a distance modulus of (m-M) = 18.40, for a linear distance of 47.9 kpc, with a statistical error of 0.15 mag or 3.4 kpc (7%).

Wolf-Rayet: based on the mean absolute magnitude of these massive stars. Applied to galaxy IC 0010, by Massey & Armandroff (1995), to obtain a distance of 0.95 Mpc.

## A.2. Standard Rulers

CO ring diameter: based on the mean absolute radius of a galaxy's inner carbon monoxide (CO) ring, with compact rings of r = ~200 pc and broad rings of r = ~750 pc. So, a CO compact ring in the galaxy Messier 82 with an apparent radius of 130 arcsec, has a distance of 3.2 Mpc (Sofue 1991).

Dwarf Galaxy Diameter: based on the absolute radii of certain kinds of dwarf galaxies surrounding giant elliptical galaxies such as Messier 87. Specifically, dwarf elliptical (dE) and dwarf spheroidal (dSph) galaxies have an effective absolute radius of ~1.0 kpc that barely varies in such galaxies over several orders of magnitude in mass. So, the apparent angular radii of these dwarf galaxies around Messier 87 at 11.46 arcseconds, gives a distance for the main galaxy of 18.0 ± 3.1 Mpc (Misgeld & Hilker 2011).

Eclipsing Binary: a hybrid method between standard rulers and standard candles, using stellar pairs orbiting one another fortuitously such that their individual masses and radii can be measured, allowing the system's absolute magnitude to be derived. Thus, the absolute visual magnitude of an eclipsing binary in the galaxy Messier 31 is $M_V$ = -5.77 (Ribas et al. 2005). So, this eclipsing binary, with an apparent visual magnitude of $m_V$ = 18.67, has a distance modulus of $(m-M)_V$ = 24.44, for a distance of 772 kpc, with a statistical error of 0.12 mag or 44 kpc (6%).

Globular Cluster Radii (GC radius): based on the mean absolute radii of globular clusters, r = 2.7 pc (Jordan et al. 2005). So, globular clusters in the galaxy Messier 87 with a mean apparent radius of r = 0.032 arcsec, have a distance of 16.4 Mpc.

Grav. Stability Gas. Disk: based on the absolute diameter at which a galaxy reaches the critical density for gravitational stability of the gaseous disk (Zasov & Bizyaev 1996). A distance to galaxy Messier 74 is obtained of 9.40 Mpc.

Gravitational Lenses (G Lens): based on the absolute distance between the multiple images of a single background galaxy that surround a gravitational lens galaxy, determined by time-delays measured between images. Thus, the apparent distance between images gives the lensing galaxy's distance. Applied to the galaxy 87GB[BWE91] 1600+4325 ABS01 by Burud et al. (2000), to obtain a distance of 1,920 Mpc.

HII Region Diameters (HII): based on the mean absolute diameter of HII regions, d = 14.9 pc (Ismail et al. 2005). So, HII regions in the galaxy Messier 101 with a mean apparent diameter of r = 4.45 arcsec, have a distance of 6.9 Mpc.

Jet Proper Motion: based on the apparent motion of individual components in parsec-scale radio jets, obtained by observation, compared with their absolute motion, obtained by Doppler measurements and corrected for the jet's angle to the line of sight. Applied to the quasar 3C 279 by Homan & Wardle (2000), to obtain an angular size distance of 1.8 ± 0.5 Gpc.

Masers: based on the absolute motion of masers orbiting at great speeds within parsecs of supermassive black holes in galaxy cores, relative to their apparent or proper motion. The absolute motion of masers orbiting within the galaxy NGC 4258 is $V_t$ = 1,075 km s$^{-1}$, or 0.001100 pc yr$^{-1}$ (Humphreys et al. 2004). So, the maser's apparent proper motion of 31.5 x 10$^{-6}$ arcsec yr$^{-1}$, gives a distance of 7.2 Mpc, with a statistical error of 0.2 Mpc (3.0%).

Orbital Mechanics (Orbital Mech.): based on the predicted orbital or absolute motion of a galaxy around another galaxy, and its observed apparent motion, giving a measure of distance. Applied by Howley et al. (2008) to the Messier 31 satellite galaxy Messier 110, to obtain a linear distance of 0.794 Mpc.

Proper Motion: based on the absolute motion of a galaxy, relative to its apparent or proper motion. Applied to galaxy Leo B by Lepine et al. (2011), to obtain a linear distance of 0.215 Mpc.

Ring Diameter: based on the apparent angular ring diameter of certain spiral galaxies with inner rings, compared to their absolute ring diameter, as determined based on other apparent properties, including morphological stage and luminosity class (Pedreros & Madore 1981). For galaxy UGC 12914, a distance modulus is obtained of (m-M) = 32.30, for a linear distance of 29.0 Mpc, with a statistical error of 0.84 mag or 13.6 Mpc (47%), assuming H = 100 km s$^{-1}$ Mpc$^{-1}$.

Type II Supernovae, Optical (SNII Optical): based on the absolute motion of the explosion's outward velocity, in units of intrinsic transverse velocity, $V_t$ (usually km s$^{-1}$), relative to the explosion's apparent or proper motion (usually arcseconds year$^{-1}$) (e.g.- Eastman, Schmidt & Kirshner 1996). So, the absolute motion of Type II SN 1979C observed in the galaxy Messier 100, based on the Expanding Photosphere Method (EPM), gives a distance of 15 Mpc, with a statistical error of 4.3 Mpc (29%). An alternative SNII Optical indicator uses the Standardized Candle Method (SCM) of Hamuy & Pinto (2002). Applied to Type II SN 2003gd in galaxy Messier 74, by Hendry et al. (2005), to obtain a distance of 9.6 Mpc, with a statistical error of 2.8 Mpc (29%).

## A.3. Secondary Methods

Brightest Cluster Galaxy (BCG): based on the fairly uniform absolute visual magnitudes of $M_V$ = -22.68 ± 0.35 found among the brightest galaxies in galaxy clusters (see Hoessel 1980). So, for example, for the brightest galaxy in the galaxy cluster Abell 0021, which is the galaxy 2MASX J00203715+2839334 and which has an apparent visual magnitude of $m_V$ = 15.13, the luminosity distance modulus can be calculated, as done by Hoessel, Gunn & Thuan (1980). The result is a luminosity distance modulus of (m-M) = 37.81, or a luminosity distance of 365 Mpc. With a redshift for the BCG in Abell 0021 of z = 0.0945, this leads to a linear distance of 333 Mpc, with a statistical error of 0.35 mag or 59 Mpc (18%).

D_n-sigma: provides standard candles based on the absolute magnitudes of elliptical and early-type galaxies, determined from the relation between the galaxy's apparent magnitude and apparent diameter (e.g., Willick et al. 1997). Applied to galaxy ESO 409- G 012, to obtain a distance modulus of (m-M) = 33.9, for a linear distance of 61 Mpc, with a statistical error of 0.40 mag or 12 Mpc (20%).

Diameter: certain galaxy's major diameters may provide secondary standard rulers based on the absolute diameter for example of only the largest, or "supergiant" spiral galaxies, estimated to be ~52 kpc (van der Kruit 1986). So, from the mean apparent diameter found for supergiant spiral galaxies in the Virgo cluster of ~9 arcmin, the Virgo cluster distance is estimated to be 20 Mpc, with a statistical error of 3 Mpc (15%).

Dwarf Ellipticals: based on the absolute magnitude of dwarf elliptical galaxies, derived from a surface-brightness/luminosity relation, and the observed apparent magnitude of these galaxies (Caldwell & Bothun 1987). Applied to Dwarf Elliptical galaxies around galaxy NGC 1316 in the Fornax galaxy cluster, to obtain a distance of 12 Mpc.

Faber-Jackson: based on the absolute magnitudes of elliptical and early-type galaxies, determined from a relation between a galaxy's apparent magnitude and velocity dispersion (Lucey 1986). Applied to galaxy NGC 4874, to obtain a distance modulus of (m-M) = 34.76, for a linear distance of 89.5 Mpc, with a statistical error of 0.12 mag or 5.1 Mpc (6%).

Fundamental Plane (FP): based on the absolute magnitudes of early-type galaxies, which depend on effective visual radius $r_e$, velocity dispersion sigma, and mean surface brightness within the effective radius $I_e$: log D = log $r_e$ - 1.24 log sigma + 0.82 log $I_e$ + 0.173 (e.g., Kelson et al. 2000). The galaxy NGC 1399 has an effective radius $r_e$ = 55.4 arcsec, a rotational velocity sigma = 301 km s$^{-1}$, and surface brightness, $I_e$ = 428.5 $L_{Sun}$ pc$^{-2}$. So, from the FP relation, its distance is 20.6 Mpc.

GC K vs. (J-K): the globular cluster K-band magnitude vs. J-band minus K-band Color-Magnitude Diagram secondary standard candle is similar to the Color-Magnitude Diagram standard candle, but applied specifically to globular clusters within a galaxy, rather than entire galaxies (Sitko 1984). Applied to galaxy Messier 31, to obtain a linear distance of 0.689 Mpc.

GeV TeV ratio: based on the absolute magnitude at which this ratio equals one, which compares energy emitted at two wavelengths, Giga-electron Volt and Tera-electron Volt (Prandini et al. 2010). Applied to galaxy 3C66A, to obtain a linear distance of 794 Mpc.

Globular Cluster Fundamental Plane (GC FP): based on the relationship among velocity dispersion, radius, and mean surface brightness for globular clusters, similar to the fundamental plane for early-type galaxies (Strader et al. 2009). Applied to globular clusters in galaxy Messier 31, to obtain a distance modulus of (m-M) = 24.57, for a linear distance of 0.820 Mpc, with a statistical error of 0.05 mag or 0.019 Mpc (2.3%).

H I + optical distribution: based on neutral Hydrogen I mass versus optical distribution or virial mass provides a secondary standard ruler that applies to extreme H I-rich galaxies, such as Michigan 160, based on the assumption that the distance-dependent ratio of neutral gas to total (virial) mass should equal one (Staveley-Smith et al. 1990). Applied to galaxy UGC 12578, to obtain a distance modulus of (m-M) = 33.11, for a linear distance of 41.8 Mpc, with a statistical error of 0.20 mag or 4.0 Mpc (10%).

Infra-Red Astronomical Satellite (IRAS): based on a reconstruction of the local galaxy density field using a model derived from the 1.2-Jy IRAS survey with peculiar velocities accounted for using linear theory (e.g., Willick et al. 1997). Applied to galaxy UGC 12897, to obtain a distance modulus of (m-M) = 35.30, for a linear distance of 115 Mpc, with a statistical error of 0.80 mag or 51 Mpc (44%).

LSB galaxies: based on the Surface Brightness Fluctuations (SBF) standard candle, which is based on the fluctuations in surface brightness arising from the mottling of the otherwise smooth light of a galaxy due to individual stars, but applied specifically to Low Surface Brightness (LSB) galaxies (Bothun et al. 1991). Applied to LSB galaxies around galaxy NGC 1316 in the Fornax galaxy cluster, to obtain a distance modulus of (m-M) = 31.25, for a linear distance of 17.8 Mpc, with a statistical error of 0.28 mag or 2.4 Mpc (14%).

Magnetic Energy: based on an extragalactic object's magnetic energy and particle energy, and calculations assuming certain relations between the two. It has been applied so far to only one gamma-ray source, HESS J1507-622 (Domainko, 2014). Depending on which theoretical possibilities are assumed, the distance is estimated to range from 0.18 Mpc to 100 Mpc, indicating that HESS J1507-622 is extragalactic.

Magnitude: based on the apparent magnitudes of certain galaxies, which may provide a secondary standard candle based on the mean absolute magnitude determined from a sample of similar galaxies with known distances. Assuming a mean absolute blue magnitude for dwarf galaxies of $M_B$ = -10.70, the dwarf galaxy DDO 155 with an apparent blue magnitude of $m_B$ = 14.5, has a distance modulus of (m-M)$_B$ = 25.2, for a distance of 1.1 Mpc (Moss & de Vaucouleurs 1986).

Mass Model: based on the absolute radii of galaxy halos, estimated from the galaxy plus halo mass as derived from rotation curves and from the expected mass density derived theoretically (Gentile et al. 2010). Applied to galaxy NGC 1560, to obtain a linear distance of 3.16 Mpc.

Radio Brightness: based on the absolute radio brightness assumed versus the apparent radio brightness observed in a galaxy (Wiklind & Henkel, 1990). Applied to galaxy NGC 0404, to obtain a distance of 10 Mpc.

Sosies: "Look Alike", or in French "Sosies", galaxies provide standard candles based on a mean absolute visual magnitude of $M_V$ = -21.3 found for spiral galaxies with similar Hubble stages, inclination angle, and light concentrations (Terry et al. 2002). So, the galaxy NGC 1365, with an apparent visual magnitude of $m_V$ = 9.63, has a distance modulus of $(m-M)_V$ = 30.96, for a distance of 15.6 Mpc. Galaxy NGC 1024, with an apparent visual magnitude of $m_V$ = 12.07 that is 2.44 mag fainter and apparently farther than NGC 1365, is also estimated to be 0.06 mag less luminous than NGC 1365, leading to a distance modulus of $(m-M)_V$ = 33.34, for a distance of 46.6 Mpc.

Tertiary: a catch-all term for various distance indicators employed by de Vaucouleurs et al. in the 1970s and 1980s, including galaxy luminosity index and rotational velocity (e.g., McCall 1989). Applied to galaxy IC 0342, to obtain a distance modulus of (m-M) = 26.32, for a linear distance of 1.84 Mpc, with a statistical error of 0.15 mag or 0.13 Mpc (7%).

Tully Estimate (Tully est): based on various parameters, including galaxy magnitudes, diameters, and group membership (Tully, Nearby Galaxies Catalog, 1988). For galaxy ESO 012- G 014, the estimated distance is 23.4 Mpc.

Tully-Fisher: introduced by Tully & Fisher (1977), based on the absolute blue magnitudes of spiral galaxies, which depend on their apparent blue magnitude, $m_B$, and their maximum rotational velocity, sigma: $M_B$ = -7.0 log sigma - 1.8 (e.g., Karachentsev et al. 2003). So, the galaxy NGC 0247 has an absolute blue magnitude of $M_B$ = -18.2, based on its rotational velocity, sigma = 222 km s$^{-1}$. With an apparent blue magnitude of $m_B$ = 9.86, NGC 0247 has a distance modulus of $(m-M)_B$ = 28.1, for a distance of 4.1 Mpc.

### A.4. Additional Information on Indicators

Here are some notes relating to Cepheids distances in particular, and to standard candle indicators in general, regarding different luminosity relations, apparent versus reddening-corrected distance, and corrections related to age or metallicity.

### A.4.1. Period–Luminosity Relation

Cepheid variable stars have absolute visual magnitudes related to the log of their periods in days

$$M_V = -2.76 \log P - 1.46$$

This is the PL relation adopted by NASA's Hubble Space Telescope Key Project On the Extragalactic Distance Scale (Freedman et al. 2001).

In the galaxy NGC 1637, the longest period Cepheid of 18 observed has a period of 54.42 days, yielding a mean absolute visual magnitude of $M_V$ = -6.25 (Leonard et al. 2003). With the star's apparent mean visual magnitude of $m_V$ = 24.19, its apparent visual distance modulus of is $(m-M)_V$ = 30.44, corresponding to a distance of 12.2 Mpc.

NGC 1637's shortest period Cepheid, with a period of 23.15 days, has a mean absolute visual magnitude of $M_V$ = -5.23. The shorter period variable's mean apparent visual magnitude is $m_V$ = 25.22, giving an apparent visual distance modulus of $(m-M)_V$ = 30.45, for a distance of 12.3 Mpc. This is in excellent agreement with the distance found from the longest-period Cepheid in the same galaxy.

### A.4.2. Apparent Distance

Nevertheless, there is in practice a significant scatter in the individual Cepheid distance moduli within a single galaxy. In the galaxy NGC 1637, for example, the average of the apparent distance moduli for all 18 Cepheids is $(m-M)_V$ = 30.76, corresponding to a distance of 14.2 Mpc. This is ~0.3 mag fainter than the distance moduli obtained from either the longest or shortest period Cepheids, and corresponds to a 15% greater distance.

### A.4.3. Reddening-corrected Distance

Scatter in individual Cepheid distance moduli is caused primarily by differential "reddening" or dimming due to differing patches of dust within target galaxies, and to a lesser extent by reddening due to foreground dust within the Milky Way, as well as differences in the intervening Intergalactic Medium. Because reddening is wavelength-dependent (greater at shorter wavelengths) the difference between distance moduli measured at two or more wavelengths can be used to estimate the extinction at any wavelength, $E_{V-I} = (m-M)_V - (m-M)_I$. For NGC 1637, with $(m-M)_{V-I}$ = 30.76 - 30.54, the extinction between V and I is $E_{V-I}$ = 0.22. Extinction, when multiplied by the ratio of total-to-selective absorption and assuming that ratio to be $R_V$ = 2.45, equals the total absorption, or dimming in magnitudes of the visual distance modulus due to dust, $A_V = R_V \times E_{V-I}$ = 0.54 in the case of NGC 1637. Note different total-to-selective absorption ratios are assumed by different authors. The correction for dimming due to dust obtained by Leonard et al. (2003) is deducted from the apparent visual distance modulus of $(m-M)_V$ = 30.76 to obtain the true, reddening-corrected, "Wesenheit" distance modulus of $(m-M)_W$ = 30.23, corresponding to a distance of 11.1 Mpc.

### A.4.4. Metallicity-corrected Distance

Cepheids formed in galaxies with higher "metal" abundance ratios (represented here by measured oxygen/hydrogen ratios), are comparatively less luminous than Cepheids formed in "younger" less evolved galaxies.

Leonard et al. (2003) apply a metallicity correction of Z = 0.12 mag, based on the difference in metal abundance between galaxy NGC 1637 and the LMC. Their final, metallicity- and reddening-corrected distance modulus is $(m-M)_Z$ = 30.34, corresponding to a distance of 11.7 Mpc.

Different corrections for reddening and age or metallicity are applied by different authors. For review see Freedman & Madore (2010).

### A.4.5. Distance Precision

Differences affecting distance estimates, whether based on Cepheid variables or other methods, include corrections for:

1. line-of-sight extinction due to foreground, target, and IGM obscuration
2. age/metallicity/colors
3. distance scale zero point
4. distance scale formula (PL or other relation)
5. photometric zero point
6. other biases; for example, the well-known Malmquist bias
7. cosmological priors; for example, the Hubble constant

All these involve systematic and statistical errors; Freedman et al. (2001) have a discussion of the errors involved in many of the methods discussed here.

FIGURES 1 to 5

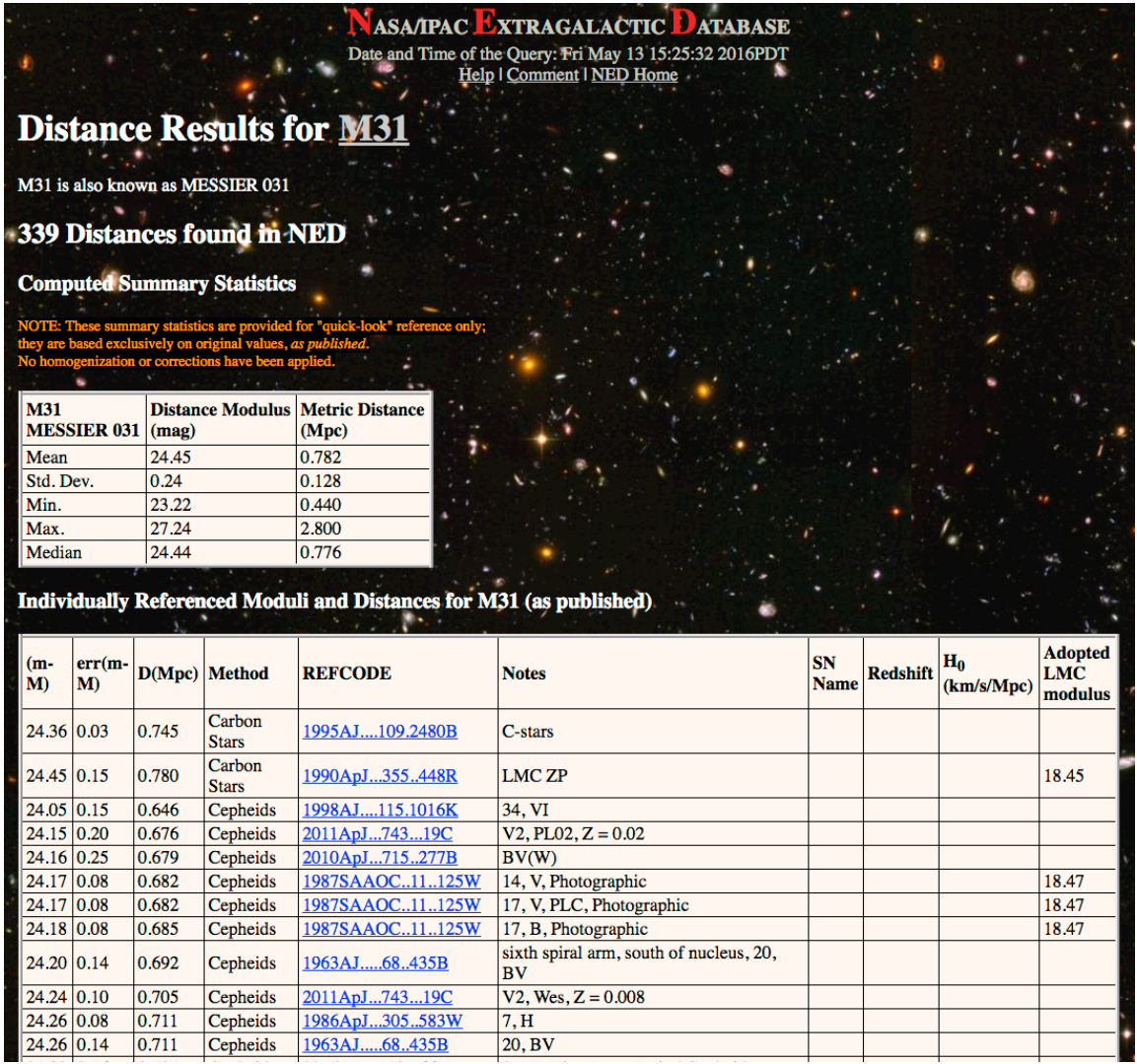

Figure 1. Screenshot of distances available for Messier 31 via the "Redshift-Independent Distances" service available on NED's main interface page. Summary statistics (mean, median, minimum, maximum, and standard deviation) are presented for "quick-look" reference, with no attempt to apply corrections, weightings or standardization. These are followed by a tabulation of all available distance estimates. The latest full tabulation of distances for M31 (339 estimates at the time of writing) is available at any time from this URL: http://ned.ipac.caltech.edu/cgi-bin/nDistance?name=M31

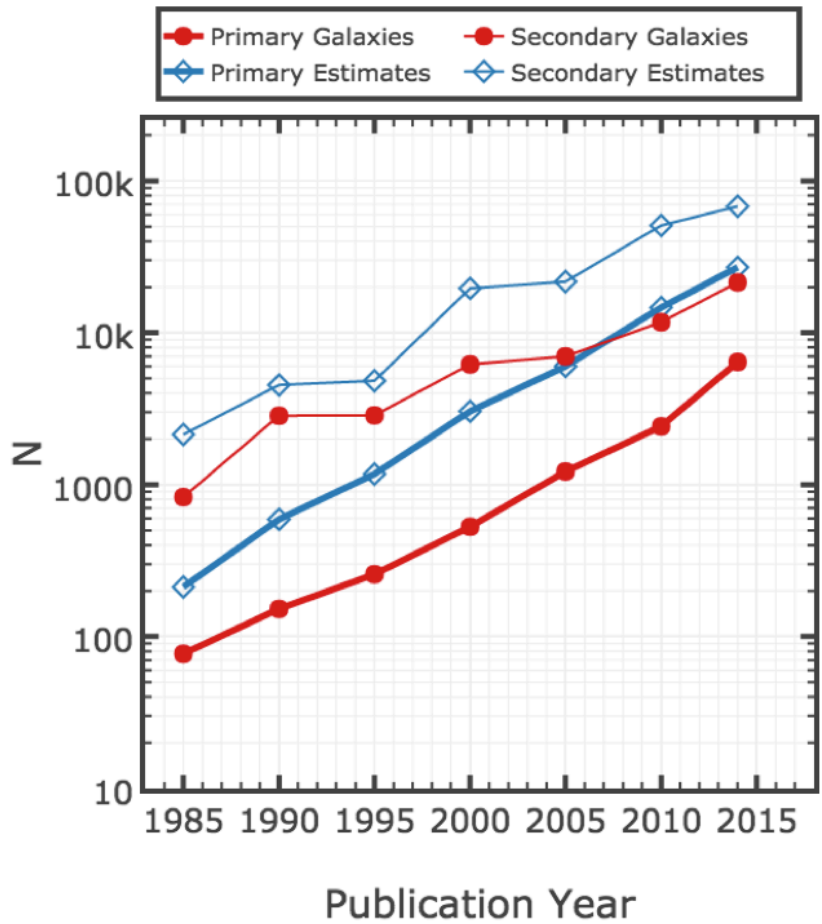

Figure 2. Growth in the number of individual redshift-independent distance estimates (blue diamonds and lines), and galaxies with such estimates (red circles and lines), is shown for primary indicators (thick lines) and secondary indicators (thin lines). Cumulative totals are plotted for the end of each five-year period, except the most recent period that is current through 2014.

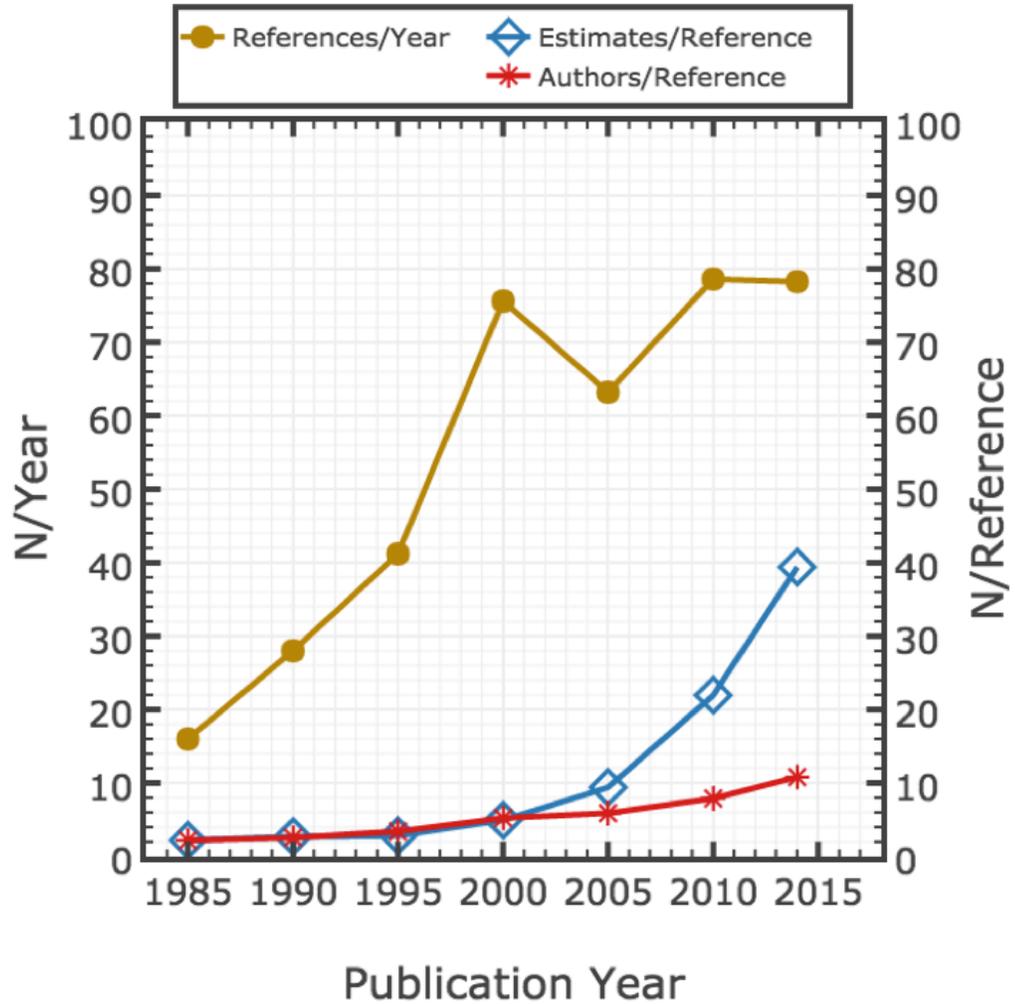

Figure 3. Growth in primary distance estimates is largely attributable to increases over time in the number of references per year (gold circles and line), the number of estimates per reference (blue diamonds and line), and the number of authors per reference (red asterisks and line). Over the period 1985 through 2014, these metrics have increased by factors of 5, 20, and 5, respectively.

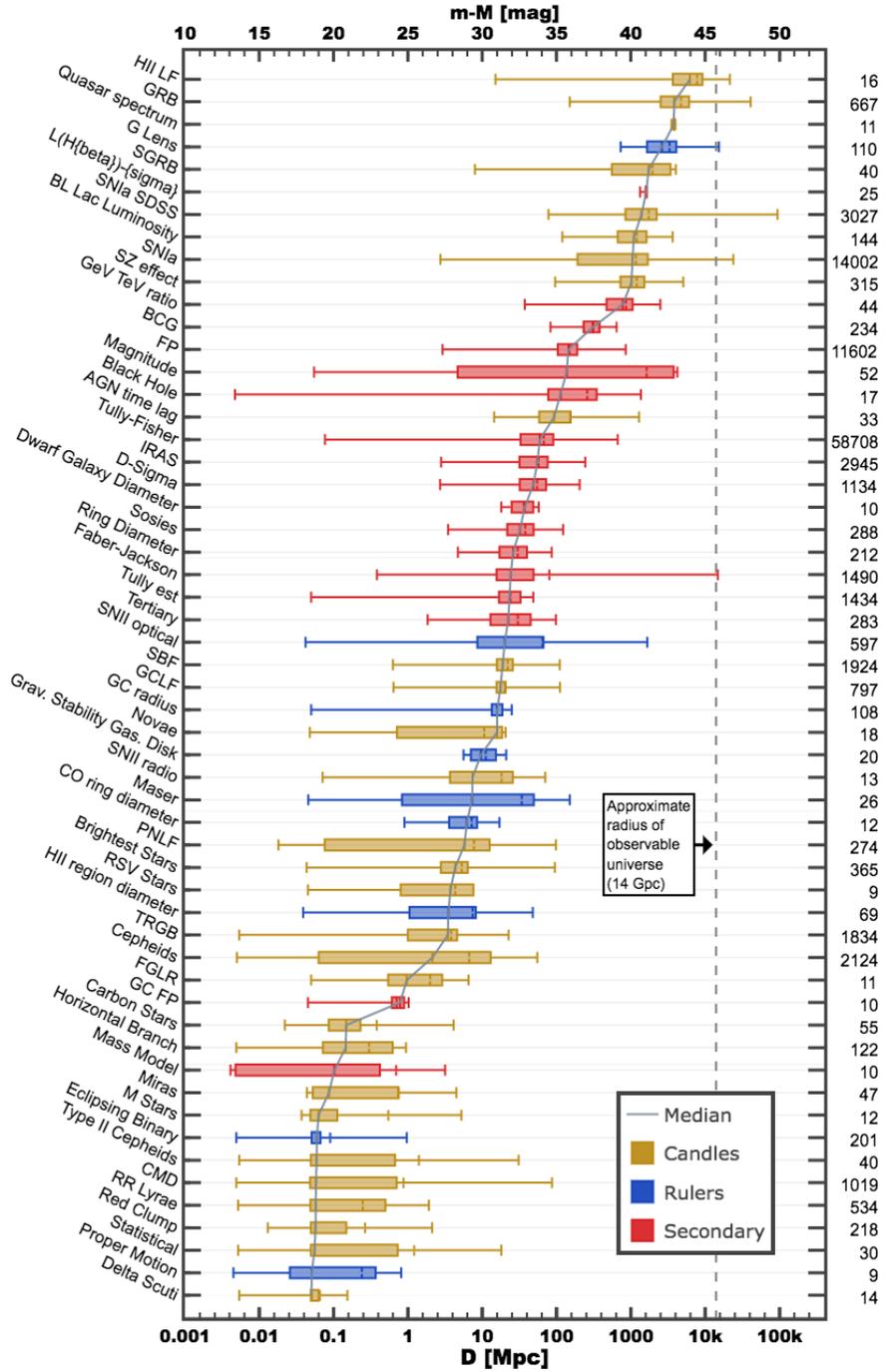

Figure 4. Comparison of redshift-independent distance indicators, shown in order of increasing median distance. For each indicator with at least nine estimates, a 'box plot' represents the distribution of the distance estimates: the left and right sides of each box represent the 25th and 75th percentiles of the distribution, and the lines extend to the minimum and maximum values. The solid curve connects the median distances of the indicators. Standard candles, standard rulers, and secondary indicators are plotted in gold, blue, and red, respectively, as shown in the legend. For each indicator labeled on the left, the corresponding number of individual distance estimates (as of 2016 July) is listed on the right.

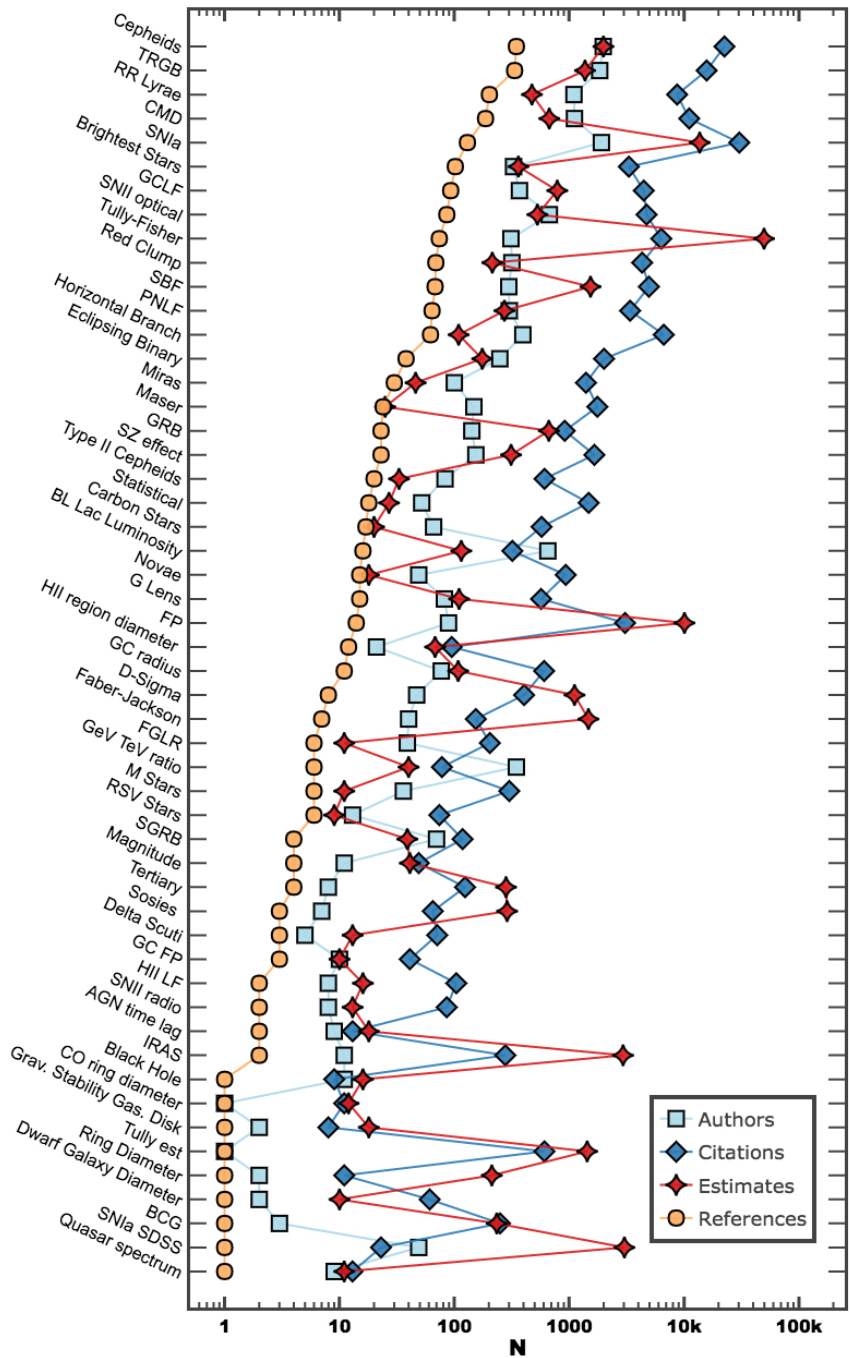

Figure 5. Research activity is shown for each distance indicator having nine or more estimates available at the time of writing. The indicators are sorted by increasing number of references, with number of authors, citations, and estimates shown by symbols indicated in the legend.

TABLES I to III

Table I. Log of major updates to NED-D

| Date | Version | Estimates | Galaxies | References |
|------|---------|-----------|----------|------------|
| 2006 | 1.0 | 3065 | 1073 | 329 |
| 2007 | 2.1 | 3716 | 1210 | 524 |
| 2008 | 2.2 | 18150 | 5049 | 526 |
| 2009 | 3.0 | 35348 | 9120 | 848 |
| 2010 | 4.0 | 36413 | 9123 | 1057 |
| 2011 | 6.0 | 40386 | 9568 | 1381 |
| 2012 | 7.0 | 60230 | 12339 | 1384 |
| 2013 | 8.0 | 67775 | 14679 | 1563 |
| 2014 | 9.0 | 70003 | 15050 | 1612 |
| 2015 | 10.0 | 85809 | 18470 | 1942 |
| 2016 | 12.0 | 107429 | 28347 | 2145 |

Note.— Only major updates are shown, out of more than 30 updates made, over the last decade. For that reason, Version 5.0, and Version 11.0 do not appear. To aid in reproducibility of future studies, older versions are available via the NED-D main page online (http://ned.ipac.caltech.edu/Library/Distances/).

Table II. Structure of the tabular version of NED-D

| Field | Label | Description |
|---|---|---|
| 1 | Exclusion Code | Exclusion Code: indicates where "" (blank) which of the distances from the master file are from peer-reviewed sources incorporated within NED, and are both interactive within NED and available via the Redshift Independent Distances query, and indicates by the letter "R" and "N" which are among the distances not yet included in NED but available in the NED-D tabulation. |
| 2 | D | Record index. |
| 3 | G | Object index. |
| 4 | Galaxy ID | NED "Preferred Object Name" for the host galaxy. |
| 5 | m-M | Distance Modulus expressed in mag. |
| 6 | err | Quoted (one-sigma) statistical (random) error on the distance modulus. |
| 7 | D (Mpc) | Metric distance (in units of Mpc). |
| 8 | Method | Distance indicator (method) used; see Table 3 for explanations of the codes. |
| 9 | REFCODE | REFCODE for the originating paper publishing the distance. Ancillary information of the various methods, such as added corrections, zero points, etc. |
| 10 | SN ID | Supernova Name: informs users when distances are based on Type Ia Supernova (SNIa), Type II Supernova optical (SNII optical), or Type II Supernova radio (SNII radio) methods, stating which SN is referenced, easing interactive comparison between NED-D and the author(s) data, which are most often presented in SN order whether by date or name, rather than order of galaxy host position, as given here. |
| 11 | redshift (z) | Redshift: appears only in cases where the distance modulus is published as a "luminosity distance modulus", as provided mostly for Type Ia supernova (SNIa), showing the target redshift used to transform each "luminosity distance modulus" given to the corresponding "metric distance", via m-M(L) = 5 x logD -5/(1+z), with D in pc. |
| 12 | Hubble const. | Hubble constant ($H_0$): appears only in cases where the $H_0$ value assumed by the author(s) differs from the default value of $H_0 = 70$ km s$^{-1}$ Mpc$^{-1}$ used here and by the Supernova Cosmology Project, the Supernova Legacy Survey and others. See for example Astier et al. (2006), who round down the value of $H_0 = 72$ km s$^{-1}$ Mpc$^{-1}$ from the NASA HST Key Project (Freedman et al. 2001). |
| 13 | Adopted LMC modulus | LMC zero point: appears only in cases where the zero point assumed by the author(s) differs from a fiduciary value of 18.50 mag. |
| 14 | Date (Yr. - 1980) | Reference Date |
| 15 | Notes | Notes, where necessary, about relevant measurement data. |

## Table III: NED-D redshift-independent extragalactic distance indicators

| Indicator | Estimates | Galaxies | Refcodes | Authors | Citations | D Min (Mpc) | D Mean (Mpc) | D Max (Mpc) | Err (mag) | Est. with Err (%) |
|---|---|---|---|---|---|---|---|---|---|---|
| Standard Candles (41) | | | | | | | | | | |
| AGB | 3 | 2 | 3 | 5 | 91 | 0.535 | 7.27 | 14.9 | 0.19 | 67 |
| AGN time lag | 18 | 18 | 2 | 9 | 13 | 14.5 | 76.1 | 146 | 0.16 | 100 |
| B Stars | 2 | 1 | 2 | 3 | 38 | 0.0460 | 0.0518 | 0.0575 | 0.25 | 100 |
| BL Lac Luminosity | 115 | 99 | 16 | 652 | 320 | 120 | 1050 | 3600 | 0.31 | 10 |
| Blue Supergiant | 2 | 2 | 1 | 1 | 5 | 0.0501 | 0.0566 | 0.0630 | 0.50 | 100 |
| Brightest Stars | 361 | 171 | 102 | 328 | 3312 | 0.0435 | 5.08 | 25.1 | 0.42 | 35 |
| Carbon Stars | 20 | 15 | 17 | 66 | 575 | 0.0310 | 0.837 | 4.11 | 0.23 | 60 |
| Cepheids | 1987 | 100 | 347 | 1980 | 22527 | 0.0355 | 6.73 | 55.0 | 0.10 | 91 |
| CMD | 671 | 136 | 187 | 1112 | 11097 | 0.0060 | 1.28 | 86.7 | 0.12 | 65 |
| Delta Scuti | 13 | 4 | 3 | 5 | 71 | 0.0492 | 0.0700 | 0.153 | 0.10 | 92 |
| FGLR | 11 | 9 | 6 | 39 | 204 | 0.0501 | 1.99 | 6.55 | 0.09 | 55 |
| GRB | 665 | 218 | 23 | 142 | 915 | 151 | 4730 | 40700 | 0.77 | 81 |
| GCLF | 788 | 206 | 93 | 369 | 4463 | 0.640 | 20.0 | 111 | 0.24 | 98 |
| GC SBF | 2 | 1 | 1 | 8 | 59 | 0.817 | 0.825 | 0.832 | 0.12 | 100 |
| HII LF | 16 | 16 | 2 | 8 | 104 | 15.1 | 7770 | 21400 | 1.24 | 100 |
| Horizontal Branch | 109 | 49 | 62 | 395 | 6668 | 0.0185 | 0.312 | 0.940 | 0.13 | 74 |
| M Stars | 11 | 7 | 6 | 36 | 301 | 0.0372 | 0.585 | 5.25 | 0.09 | 45 |
| Miras | 46 | 14 | 30 | 100 | 1399 | 0.0439 | 0.744 | 4.49 | 0.18 | 72 |
| Novae | 18 | 7 | 15 | 49 | 933 | 0.0479 | 10.6 | 20.6 | 0.36 | 78 |
| OB Stars | 5 | 2 | 5 | 10 | 376 | 0.0457 | 0.0566 | 0.0661 | 0.28 | 80 |
| PNLF | 273 | 77 | 64 | 301 | 3397 | 0.0181 | 7.73 | 97.7 | 0.18 | 71 |
| PAGB Stars | 2 | 2 | 1 | 2 | 1 | 0.692 | 0.753 | 0.813 | 0.00 | 0 |
| Quasar spectrum | 11 | 11 | 1 | 9 | 13 | 3520 | 3775 | 3970 | 0.00 | 0 |
| Red Clump | 214 | 27 | 69 | 317 | 4323 | 0.0130 | 0.269 | 2.11 | 0.08 | 75 |
| RSV Stars | 9 | 6 | 6 | 13 | 74 | 0.0453 | 4.32 | 7.59 | 0.19 | 100 |
| RV Stars | 5 | 1 | 1 | 2 | 38 | 0.0605 | 0.0613 | 0.0619 | 0.05 | 100 |
| RR Lyrae | 474 | 50 | 202 | 1101 | 8703 | 0.0060 | 0.248 | 1.91 | 0.12 | 71 |
| S Doradus Stars | 5 | 5 | 1 | 1 | 58 | 0.759 | 2.93 | 5.25 | 0.00 | 0 |
| SGRB | 39 | 35 | 4 | 70 | 118 | 8.00 | 1900 | 4030 | 0.94 | 21 |
| Subdwarf fitting | 1 | 1 | 1 | 4 | 298 | 0.0535 | 0.0535 | 0.0535 | 0.12 | 100 |
| SZ effect | 312 | 49 | 23 | 154 | 1658 | 96.0 | 1198 | 5070 | 0.55 | 43 |
| SNIa | 13700 | 3130 | 130 | 1908 | 30282 | 2.73 | 1190 | 23900 | 0.21 | 97 |
| SNIa SDSS | 3027 | 1772 | 1 | 49 | 23 | 77.6 | 1740 | 94300 | 0.30 | 100 |
| SNII radio | 13 | 13 | 2 | 8 | 86 | 0.0710 | 18.2 | 70.7 | 0.36 | 92 |
| SBF | 1534 | 545 | 68 | 298 | 4948 | 0.637 | 22.1 | 110 | 0.20 | 99 |
| SX Phe Stars | 2 | 2 | 2 | 2 | 57 | 0.0279 | 0.0640 | 0.100 | 0.05 | 50 |
| TRGB | 1374 | 352 | 335 | 1845 | 15730 | 0.0071 | 3.34 | 20.0 | 0.12 | 82 |
| Type II Cepheids | 33 | 15 | 20 | 83 | 610 | 0.0472 | 1.69 | 31.0 | 0.14 | 85 |
| White Dwarfs | 1 | 1 | 1 | 4 | 298 | 0.0479 | 0.0479 | 0.0479 | 0.15 | 100 |
| Wolf-Rayet | 3 | 1 | 3 | 7 | 128 | 0.870 | 1.11 | 1.50 | 0.34 | 33 |
| Statistical | 27 | 13 | 18 | 52 | 1478 | 0.0481 | 1.35 | 18.0 | 0.09 | 96 |
| Standard Rulers (10) | | | | | | | | | | |
| CO ring diameter | 12 | 12 | 1 | 1 | 11 | 0.900 | 7.22 | 17.0 | 0.00 | 0 |
| Eclipsing Binary | 175 | 5 | 38 | 249 | 2014 | 0.0209 | 0.0968 | 0.964 | 0.08 | 70 |
| GC radius | 108 | 107 | 11 | 77 | 605 | 0.0501 | 15.7 | 24.9 | 0.17 | 94 |
| G Lens | 110 | 49 | 15 | 82 | 570 | 730 | 3310 | 15300 | 0.58 | 72 |
| Grav. Stability Gas. Disk | 18 | 18 | 1 | 2 | 8 | 5.60 | 10.2 | 18.1 | 0.00 | 0 |
| HII region diameter | 68 | 42 | 12 | 21 | 95 | 0.0391 | 7.50 | 47.7 | 0.14 | 76 |
| Jet Proper Motion | 1 | 1 | 1 | 2 | 10 | 1800 | 1800 | 1800 | 0.53 | 100 |
| Maser | 25 | 9 | 24 | 148 | 1762 | 0.0457 | 31.1 | 151 | 0.29 | 80 |
| Proper Motion | 7 | 4 | 3 | 9 | 48 | 0.0328 | 0.308 | 0.809 | 0.50 | 43 |
| SNII optical | 529 | 107 | 86 | 672 | 4722 | 0.0421 | 70.7 | 1660 | 0.27 | 96 |
| Secondary Indicators (24) | | | | | | | | | | |
| BCG | 234 | 234 | 1 | 3 | 250 | 82.9 | 309 | 643 | 0.35 | 100 |
| Black Hole | 16 | 16 | 1 | 11 | 9 | 17.9 | 274 | 1360 | 0.71 | 100 |
| D_n-sigma | 1114 | 547 | 8 | 47 | 405 | 2.70 | 54.6 | 205 | 0.40 | 99 |
| Diameter | 3 | 2 | 2 | 2 | 130 | 20.0 | 117 | 214 | 0.41 | 67 |
| Dwarf Ellipticals | 1 | 1 | 1 | 2 | 148 | 12.0 | 12.0 | 12.0 | 0.00 | 0 |

| Method | | | | | | | | | | |
|---|---|---|---|---|---|---|---|---|---|---|
| Dwarf Galaxy Diameter | 10 | 5 | 1 | 2 | 61 | 18.0 | 36.4 | 57.6 | 0.68 | 100 |
| Faber-Jackson | 1472 | 438 | 7 | 40 | 154 | 0.382 | 80.3 | 14800 | 0.59 | 95 |
| FP | 10093 | 9780 | 14 | 89 | 3071 | 2.92 | 154 | 507 | 0.56 | 99 |
| GC K vs. (J-K) | 1 | 1 | 1 | 1 | 19 | 0.689 | 0.689 | 0.689 | 0.00 | 0 |
| GeV TeV ratio | 40 | 22 | 6 | 347 | 78 | 37.3 | 759 | 2110 | 0.52 | 58 |
| GC FP | 10 | 2 | 3 | 10 | 41 | 0.0453 | 0.705 | 1.02 | 0.23 | 30 |
| H I + optical distribution | 1 | 1 | 1 | 5 | 26 | 41.8 | 41.8 | 41.8 | 0.20 | 100 |
| IRAS | 2945 | 2438 | 2 | 11 | 278 | 2.80 | 57.2 | 244 | 0.80 | 100 |
| LSB galaxies | 6 | 6 | 1 | 3 | 89 | 13.6 | 19.1 | 25.2 | 0.28 | 17 |
| Magnetic energy | 3 | 1 | 1 | 1 | 2 | 0.180 | 4.00 | 100 | 6.50 | 33 |
| Magnitude | 41 | 41 | 4 | 11 | 49 | 0.0547 | 2040 | 4210 | 0.15 | 95 |
| Mass Model | 4 | 2 | 3 | 6 | 10 | 0.0950 | 1.58 | 3.16 | 1.23 | 50 |
| Orbital Mech. | 4 | 4 | 4 | 13 | 59 | 0.0470 | 0.708 | 1.75 | 0.26 | 75 |
| Radio Brightness | 1 | 1 | 1 | 2 | 42 | 10.0 | 10.0 | 10.0 | 0.00 | 0 |
| Ring Diameter | 212 | 165 | 1 | 2 | 11 | 4.70 | 30.0 | 86.0 | 0.80 | 100 |
| Sosies | 288 | 288 | 3 | 7 | 65 | 3.47 | 38.0 | 123 | 0.25 | 100 |
| Tertiary | 282 | 281 | 4 | 8 | 124 | 1.84 | 30.2 | 97.7 | 0.39 | 100 |
| Tully est | 1434 | 1431 | 1 | 1 | 609 | 0.0499 | 23.9 | 48.6 | 0.80 | 100 |
| Tully-Fisher | 49768 | 11143 | 74 | 311 | 6338 | 0.0766 | 66.1 | 596 | 0.43 | 85 |

Note.— Err (mag) and Est. with Err (%) represent mean of errors published, given in units of magnitude (m-M), and percentage of number of estimates in total for which error is available.